
\font\tenmib=cmmib10
\font\sevenmib=cmmib10 scaled 800
\font\titolo=cmbx12
\font\titolone=cmbx10 scaled\magstep 2
\font\cs=cmcsc10

\font\ninerm=cmr9
\font\ottorm=cmr8
\textfont5=\tenmib
\scriptfont5=\sevenmib
\scriptscriptfont5=\fivei

\font\msytw=msbm10 scaled\magstep1

\font\msytwww=msbm7 scaled\magstep1
\font\indbf=cmbx10 scaled\magstep2

%
%
%
%
%
%
%

\global\newcount\numsec\global\newcount\numapp
\global\newcount\numfor\global\newcount\numfig\global\newcount\numsub
\numsec=0\numapp=0\numfig=1
\def\veroparagrafo{\number\numsec}\def\veraformula{\number\numfor}
\def\veraappendice{\number\numapp}\def\verasub{\number\numsub}
\def\verafigura{\number\numfig}

\def\section(#1,#2){\advance\numsec by 1\numfor=1\numsub=1%
\SIA p,#1,{\veroparagrafo} %
\write15{\string\Fp (#1){\secc(#1)}}%
\write16{ sec. #1 ==> \secc(#1)  }%
\hbox to \hsize{\titolo\hfill \number\numsec. #2\hfill%
\expandafter{\alato(sec. #1)}}\*}

\def\appendix(#1,#2){\advance\numapp by 1\numfor=1\numsub=1%
\SIA p,#1,{A\veraappendice} %
\write15{\string\Fp (#1){\secc(#1)}}%
\write16{ app. #1 ==> \secc(#1)  }%
\hbox to \hsize{\titolo\hfill Appendix A\number\numapp. #2\hfill%
\expandafter{\alato(app. #1)}}\*}

\def\senondefinito#1{\expandafter\ifx\csname#1\endcsname\relax}

\def\SIA #1,#2,#3 {\senondefinito{#1#2}%
\expandafter\xdef\csname #1#2\endcsname{#3}\else
\write16{???? ma #1#2 e' gia' stato definito !!!!} \fi}

\def \Fe(#1)#2{\SIA fe,#1,#2 }
\def \Fp(#1)#2{\SIA fp,#1,#2 }
\def \Fg(#1)#2{\SIA fg,#1,#2 }

\def\etichetta(#1){(\veroparagrafo.\veraformula)%
\SIA e,#1,(\veroparagrafo.\veraformula) %
\global\advance\numfor by 1%
\write15{\string\Fe (#1){\equ(#1)}}%
\write16{ EQ #1 ==> \equ(#1)  }}

\def\etichettaa(#1){(A\veraappendice.\veraformula)%
\SIA e,#1,(A\veraappendice.\veraformula) %
\global\advance\numfor by 1%
\write15{\string\Fe (#1){\equ(#1)}}%
\write16{ EQ #1 ==> \equ(#1) }}

\def\getichetta(#1){Fig. \verafigura%
\SIA g,#1,{\verafigura} %
\global\advance\numfig by 1%
\write15{\string\Fg (#1){\graf(#1)}}%
\write16{ Fig. #1 ==> \graf(#1) }}

\def\etichettap(#1){\veroparagrafo.\verasub%
\SIA p,#1,{\veroparagrafo.\verasub} %
\global\advance\numsub by 1%
\write15{\string\Fp (#1){\secc(#1)}}%
\write16{ par #1 ==> \secc(#1)  }}

\def\etichettapa(#1){A\veraappendice.\verasub%
\SIA p,#1,{A\veraappendice.\verasub} %
\global\advance\numsub by 1%
\write15{\string\Fp (#1){\secc(#1)}}%
\write16{ par #1 ==> \secc(#1)  }}

\def\Eq(#1){\eqno{\etichetta(#1)\alato(#1)}}
\def\eq(#1){\etichetta(#1)\alato(#1)}
\def\Eqa(#1){\eqno{\etichettaa(#1)\alato(#1)}}
\def\eqa(#1){\etichettaa(#1)\alato(#1)}
\def\eqg(#1){\getichetta(#1)\alato(fig. #1)}
\def\sub(#1){\0\palato(p. #1){\bf \etichettap(#1).}}
\def\asub(#1){\0\palato(p. #1){\bf \etichettapa(#1).}}

\def\equv(#1){\senondefinito{fe#1}$\clubsuit$#1%
\write16{eq. #1 non e' (ancora) definita}%
\else\csname fe#1\endcsname\fi}
\def\grafv(#1){\senondefinito{fg#1}$\clubsuit$#1%
\write16{fig. #1 non e' (ancora) definito}%
\else\csname fg#1\endcsname\fi}
\def\secv(#1){\senondefinito{fp#1}$\clubsuit$#1%
\write16{par. #1 non e' (ancora) definito}%
\else\csname fp#1\endcsname\fi}

\def\equ(#1){\senondefinito{e#1}\equv(#1)\else\csname e#1\endcsname\fi}
\def\graf(#1){\senondefinito{g#1}\grafv(#1)\else\csname g#1\endcsname\fi}
\def\secc(#1){\senondefinito{p#1}\secv(#1)\else\csname p#1\endcsname\fi}
\def\sec(#1){{\S\secc(#1)}}

\def\BOZZA{
\def\alato(##1){\rlap{\kern-\hsize\kern-1.2truecm{$\scriptstyle##1$}}}
\def\palato(##1){\rlap{\kern-1.2truecm{$\scriptstyle##1$}}}
}

\def\alato(#1){}
\def\galato(#1){}
\def\palato(#1){}


{\count255=\time\divide\count255 by 60 \xdef\hourmin{\number\count255}
        \multiply\count255 by-60\advance\count255 by\time
   \xdef\hourmin{\hourmin:\ifnum\count255<10 0\fi\the\count255}}

\def\oramin{\hourmin }

\def\data{\number\day/\ifcase\month\or gennaio \or febbraio \or marzo \or
aprile \or maggio \or giugno \or luglio \or agosto \or settembre
\or ottobre \or novembre \or dicembre \fi/\number\year;\ \oramin}
\setbox200\hbox{$\scriptscriptstyle \data $}
\footline={\rlap{\hbox{\copy200}}\tenrm\hss \number\pageno\hss}


\let\a=\alpha \let\b=\beta       \let\d=\delta  \let\e=\varepsilon
     \let\th=\vartheta \let\k=\kappa   \let\l=\lambda
\let\m=\mu                \let\p=\pi      \let\r=\rho
\let\s=\sigma \let\t=\tau   \let\f=\varphi     \let\c=\chi
   \let\o=\omega 
 \let\D=\Delta     \let\L=\Lambda  
     \let\F=\Phi

\def\\{\hfill\break} \let\==\equiv

\let\io=\infty 

\let\0=\noindent \def\pagina{{\vfill\eject}}

\def\ie{\hbox{\it i.e.\ }}
\let\dpr=\partial 

\def\tende#1{\,\vtop{\ialign{##\crcr\rightarrowfill\crcr
 \noalign{\kern-1pt\nointerlineskip}
 \hskip3.pt${\scriptstyle #1}$\hskip3.pt\crcr}}\,}
\def\otto{\,{\kern-1.truept\leftarrow\kern-5.truept\to\kern-1.truept}\,}
\def\fra#1#2{{#1\over#2}}

\def\VV{{\cal V}}
\def\CC{{\cal C}}\def\FF{{\cal F}}
\def\TT{{\cal T}}\def\NN{{\cal N}}\def\BB{{\cal B}}

\def\DD{{\cal D}}\def\AA{{\cal A}}

\def\T#1{{#1_{\kern-3pt\lower7pt\hbox{$\widetilde{}$}}\kern3pt}}
\def\VVV#1{{\underline #1}_{\kern-3pt
\lower7pt\hbox{$\widetilde{}$}}\kern3pt\,}
\def\W#1{#1_{\kern-3pt\lower7.5pt\hbox{$\widetilde{}$}}\kern2pt\,}
\def\Im{{\rm Im}\,}

\def\mod{{\rm mod}\,}  \def\sign{{\rm sign}\,}
\def\indica{\leaders \hbox to 0.5cm{\hss.\hss}\hfill}
\def\guida{\leaders\hbox to 1em{\hss.\hss}\hfill}
\mathchardef\oo= "0521

\def\pp{{\bf p}}\def\xx{{\bf x}}
\def\yy{{\bf y}}\def\kk{{\bf k}}

\def\qed{\raise1pt\hbox{\vrule height5pt width5pt depth0pt}}
\def\hf#1{{\hat \f_{#1}}}
 \def\tg#1{{\tilde g_{#1}}}
\def\bq{{\bar q}} \def\Val{{\rm Val}}
\def\indic{\hbox{\raise-2pt \hbox{\indbf 1}}}

\def\RRR{\hbox{\msytw R}}

\def\NNN{\hbox{\msytw N}} 
 \def\ZZZ{\hbox{\msytw Z}}

\def\ttt{\hbox{\msytwww T}}

\newcount\mgnf  
\mgnf=0

\ifnum\mgnf=0
\def\openone{\leavevmode\hbox{\ninerm 1\kern-3.3pt\tenrm1}}%
\def\*{\vglue0.3truecm}\fi
\ifnum\mgnf=1
\def\openone{\leavevmode\hbox{\ninerm 1\kern-3.63pt\tenrm1}}%
\def\*{\vglue0.5truecm}\fi

\ifnum\mgnf=0
   \magnification=\magstep0
   \hsize=14truecm\vsize=24.truecm
   \parindent=0.3cm\baselineskip=0.45cm\fi
\ifnum\mgnf=1
   \magnification=\magstep1\hoffset=0.truecm
   \hsize=14truecm\vsize=24.truecm
   \baselineskip=18truept plus0.1pt minus0.1pt \parindent=0.9truecm
   \lineskip=0.5truecm\lineskiplimit=0.1pt      \parskip=0.1pt plus1pt\fi

\openin14=\jobname.aux \ifeof14 \relax \else
\input \jobname.aux \closein14 \fi
\openout15=\jobname.aux

\null\vskip1.truecm
\def\II{{\cal I}}

\null\vskip2.truecm
\centerline{\titolone Peierls instability for the Holstein model}
\vskip.2truecm
\centerline{\titolone with rational density}
\vskip1.truecm \centerline{{\titolo G. Benfatto}\footnote{${}^\ast$}{\ottorm
Supported by MURST, Italy.}}
\centerline{Dipartimento di Matematica,
Universit\`a di Roma ``Tor Vergata''}
\centerline{Via della Ricerca Scientifica, I-00133, Roma}
\vskip.2truecm
\centerline{{\titolo G. Gentile}${}^\ast$}
\centerline{Dipartimento di Matematica, Universit\`a di Roma Tre}
\centerline{Largo San Leonardo Murialdo 1, I-00146 Roma}
\vskip.2truecm
\centerline{\titolo V. Mastropietro${}^\ast$}
\centerline{Dipartimento di Matematica, Universit\`a di Roma ``Tor Vergata''}
\centerline{Via della Ricerca Scientifica, I-00133, Roma}
\vskip1.truecm

\line{\vtop{
\line{\hskip1.5truecm\vbox{\advance \hsize by -3.1 truecm
\0{\cs Abstract.}
{\it We consider the static Holstein model,
describing a chain of Fermions interacting with a classical phonon
field, when the interaction is weak and the
density is a rational number.
We show that the energy of the system, as a function of the phonon field,
has two stationary points, defined up to a lattice translation,
which are local minima in the space of fields
periodic with period equal to the inverse of the density.}}
\hfill} }}

\pagina

\section(1,Introduction)

\sub(1.1) The {\sl Holstein model}
[P,H] was introduced to represent the interaction of
electrons with optical phonons in a crystal.
In the original model the phonons  are represented in
terms of quantum oscillators but the difficulty of understanding
such a fully quantum model has led to a modification of it, called
{\sl static Holstein model} (or {\sl adiabatic Holstein model}),
in which the phonons are classical oscillators.
This corresponds to neglect the vibrational kinetic energy of the phonons,
an approximation which can be justified in physical models
as the atom mass is much larger than the electron mass.

The Hamiltonian of the model, if we neglect all internal degrees of freedom
(the spin, for example, which play no role at zero external magnetic field)
is given by
$$\eqalign{
H&\=H_L^{\rm el}+ {1\over 2} \sum_{x\in\L} \f_x^2 \cr
&=\sum_{x,y\in\L}
t_{xy}\,\psi^+_{x}\psi^-_{y} - \mu\sum_{x\in\L}
\psi^+_{x}\psi^-_{x}
-\l\sum_{x\in\L} \f_x\psi^+_{x}\psi^-_{x}
+ {1\over 2} \sum_{x\in\L} \f_x^2 \; ,\cr} \Eq(1.1) $$
where $x,y$ are points on the one-dimensional lattice $\L$
with unit spacing, length $L$ and periodic boundary conditions; we shall
identify $\L$ with $\{x\in\ZZZ:\ -[L/2]\le x \le [(L-1)/2]\}$.
Moreover the matrix $t_{xy}$ is
defined as $t_{xy}=\d_{x,y}-(1/2)[\d_{x,y+1}+\d_{x,y-1}]$,
where $\d_{x,y}$ is the Kronecker delta,
$\m$ is the chemical potential and $\l$ is the interaction strength.
The fields $\psi_x^{\pm}$ are creation ($+$) and annihilation ($-$)
fermionic fields, satisfying periodic boundary conditions:
$\psi_x^{\pm}=\psi_{x+L}^{\pm}$. We define also
$\psi_{\xx}^{\pm}=e^{tH}\psi_x^{\pm}e^{-Ht}$, with $\xx=(x,t)$,
$-\b/2\le t \le \b/2$ for some $\b>0$; on $t$ antiperiodic boundary
conditions are imposed. The potential $\f_x$
is a real function representing the classical phonon field.

At finite $L$, the fermionic Fock space is finite dimensional, hence there is
a minimum eigenvalue $E^{\rm el}_L(\f,\m)$ of the operator $H_L^{\rm el}$,
for each given phonon field $\f$ and each value of $\m$;
let $\r_L(\f,\m)$ be the corresponding fermionic density.
The aim is to minimize the functional
$$ F_L(\f,\m)=E^{\rm el}_L(\f,\m)+ {1\over 2} \sum_{x\in\L} \f_x^2 \; ,
\Eq(1.2) $$
subject to the condition
$$ \r_L(\f,\m) = \r_L \; ,\Eq(1.3)$$
where $\r_L$ is a fixed value of the density, converging for $L\to\io$,
say to $\r$.


The model \equ(1.1) can be considered as an approximation
of a  {\it more realistic} continuous model containing also
the interaction with a fixed external periodic
potential of period one. Then the discreteness is not
a pure mathematical artifice, but it has a precise physical
interpretation: the properties of the two models are
expected to be the same, and we think that this could
be easily proven along the lines of the present paper.

\*

\sub(1.2) It is generally believed that, as a consequence
of Peierls instability argument, [P,F], there is a
field $\f^{(0)}$, uniquely defined up to a spatial translation, which
minimizes \equ(1.2), \equ(1.3), in the limit
$L\to\io$, and is a function of the form $\bar\f(2\pi\r x)$,
where $\bar\f(u)$ is a $2\pi$-periodic function.
This is physically interpreted by saying that one-dimensional
metals are unstable at low temperature, in the sense that they can lower
their energy through a periodic distortion of the ``physical
lattice'' with period $1/\r$ (in the continuous version of the model, since
$1/\r$ is not an integer in general): such a distortion is called a
{\it charge density wave}, as the physical lattice and the electronic charge
density form a new periodic structure with period bigger
than the original lattice period.

The argument in [P,F] is quite simple: the
periodic potential $\bar\f(2\pi\r x)$ opens a gap in the electronic
dispersion relation in correspondence of the Fermi momentum,
and a trivial computation using degenerate perturbation theory
shows that the elastic energy increase
is less than the fermionic energy decrease.
However, see [LRA], in this argument one does not take into account
the effects due to the discreteness of the lattice, in particular
the fact that the momenta are conserved modulo $2\pi$
({\sl Umklapp}). Neglecting the discreteness of the lattice
one loses the difference between commensurate or
incommensurate charge density wave (\ie rational or irrational $\r$) in
the infinite volume limit, whose properties are supposed to be
different, especially concerning the conductivity [F,LRA].

Note also that, even if the argument in [P,F]
is perturbative, Peierls instability is expected to arise
also for large interaction strength, [AL].

An exact result, [KL,LM], makes rigorous the
theory of Peierls instability
for the model \equ(1.1) in the case $\r=\r_L=1/2$
({\sl half filled band case}), for any value of $\l$.
In fact, in this case it has been proved
that there is a global minimum of $F_L(\f)$
of the form $\e(\l) (-1)^x$, where $\e(\l)$
is a suitable function of $\l$. This means that the periodicity
of the ground state phonon field is $2$ (recall
that in our units $1$ is just the lattice spacing):
this phenomenon is called {\sl dimerization}.
The proof heavily relies on symmetry
properties which hold only in the half filled case.

In [AAR,BM] Peierls instability for the Holstein model
is proven assuming $\l$ {\it large enough}:
in that case the Fermions are almost classical particles and
the quantum effects are treated as perturbations.
The results hold for the commensurate or incommensurate case;
in particular in the incommensurate case the function $\bar\f(u)$, related
to the minimizing field through the relation
$\f_x=\bar\f(2\pi\r x)$, has infinite many discontinuities.
On the contrary, in the small $\l$ case, according to
numerical results, $\bar\f(u)$
has been conjectured to be an analytic function of its argument,
both for the commensurate and incommensurate cases, [AAR].

In this paper we study the case of small $\l$
and {\it any} rational density, for which there are, to our knowledge,
no results in the literature besides the simulations in [AAR].
Analytical results in the small $\l$ case
can be found for a related model, the {\sl Falikov-Kimball model},
described by a Hamiltonian of the form \equ(1.1),
in which the continuous $\f_x$ is replaced by a discrete function
taking only the values 0 or 1; see [FGM].

Let $\r=P/Q$, with $P,Q$ relative prime integers,
and let $L=L_i\=iQ$, $i\in\NNN$; we shall prove that,
if $\l$ is small enough, there are two stationary points
$\f^{(\pm,i)}$ of $F_{L_i}(\f,\m)$, defined up to a lattice translation,
satisfying \equ(1.3) with $\r_L=\r$.
These stationary points are periodic functions on $L_i$ of period $Q$
(the smallest multiple of $1/\r$ which is an integer, hence a multiple
of the unit lattice spacing),
converging for $i\to\io$. Moreover, if we restrict
$F_{L_i}(\f,\m)$ to functions such that $\f_x=\f_{x+Q}$, $\f^{(\pm,i)}$ are
local minima in the norm $||\f||=\sup_{x\in L}|\f_x|$, uniformly in $i$.

The presence of the lattice has the effect that
we need the smaller $\l$ the bigger $Q$ is, see \equ(1.16).
In particular we are not able to draw conclusions about
the incommensurate case neither we know if this is a technical
limitation or there is some physical reason behind it, so that we can
not draw any conclusions about the analyticity conjecture in [AAR].

\*

\sub(1.4) Let ${\bf h}_{xy} = t_{xy} -\l\f_x\d_{xy}$ be the one-particle
Hamiltonian and $e_1(\f)\le e_2(\f) \le\ldots\le e_L(\f)$ its eigenvalues.
We have
$$E^{\rm el}_L(\f,\m) = \sum_{n:e_n(\f)\le\m} [e_n(\f)-\m] =
{\rm Tr}([{\bf h} -\m]{\bf P}_\m)\; ,\Eq(1.4)$$
where ${\bf P}_\m$ is the projector on the subspace spanned by the eigenvectors
of ${\bf h}$ with eigenvalues $\le \m$. As it is well known (see, for
example [BM]), $E^{\rm el}_L(\f,\m)$ is a differentiable function of $\f$ and,
since ${\rm Tr}({\bf h}\dpr{\bf P}_\m/\dpr\f_x)=0$,
$${\dpr\over \dpr\f_x} E^{\rm el}_L(\f,\m) = {\rm Tr}\Big(\Big[
{\dpr{\bf h}\over
\dpr\f_x}\Big] {\bf P}_\m\Big) = -\l\r_x(\f,\m) \; , \Eq(1.5)$$
where $\r_x(\f,\m)=({\bf P}_\m)_{xx}$ is the density of the electrons in the
point $x$.

Let us now suppose that $\m$ is not equal to any eigenvalue of ${\bf h}$. In
this case, given $\tilde\f$, also $\r_L(\f,\m)$ is differentiable in a
neighborhood small enough of $\tilde\f$ (so small that $e_n(\f)-\m$ stays
different from zero, for any $n$, see again [BM]) and
$\dpr\r_L(\f,\m)/\dpr\f_x = 0$. Hence, a local minimum of
\equ(1.2) satisfying \equ(1.3) must satisfy the conditions
$$\eqalign{
\f_x &= \l\r_x(\f,\m)\; ,\cr
\r_L\ &= {1\over L}\sum_x \r_x(\f,\m)\; ,\cr}\Eq(1.6)$$
$$M_{xy}\ \= \d_{xy} - \l {\dpr\over \dpr\f_x} \r_y(\f,\m) \quad \hbox{is
positive definite}\; .\Eq(1.7)$$

Note that, given $\tilde\f$, the previous condition on $\m$ can be in general
satisfied only if $||\f-\tilde\f||$ is of order $1/L$, so that a
solution of \equ(1.6) defines in general a local minimum only in a
neighborhood of size $1/L$. It follows that the only solutions which are
interesting in the limit $L\to\io$ are those associated with a gap of ${\bf h}$
around $\m$, whose size is independent of $L$.

\*

\sub(1.5) If $\f$ is a solution of \equ(1.6), it must satisfy the
condition $\hat\f_0=L^{-1}\sum_x \f_x$ $=\l\r_L$. On the other hand, if we
define $\c_x=\f_x-\hat\f_0$, we can see immediately that $\r_L(\f,\m)=
\r_L(\c,\m+\l\hat\f_0)$. It follows that we can restrict our search of
local minima of \equ(1.2) to fields $\f$ with zero mean, satisfying the
conditions
$$\eqalign{
\f_x &= \l(\r_x(\f,\m)-\r_L)\; ,\cr
\r_L\ &= {1\over L}\sum_x \r_x(\f,\m)\; ,\cr}\Eq(1.8)$$
and condition \equ(1.7).

Of course, if the field $\f_x$ satisfies \equ(1.8), the same is true for
the translated field $\f_{x+n}$, for any integer $n$. On the other hand,
one
expects that the solutions of \equ(1.8) are even with respect to some
point
of $\L$; hence we can eliminate the trivial source of non-uniqueness
described above by imposing the further
condition $\f_x=\f_{-x}$. We shall then consider only fields of the form
$$\f_x=\sum_{n=-[L/2]}^{[(L-1)/2]} \hat\f_n'
e^{i2n\p x \over L} \; , \qquad \hf{-n}'=\hf{n}'\in\RRR\; ,\qquad
\hat\f_0 = 0\; .\Eq(1.9) $$

As we said in \sec(1.2), we want to consider the case of rational density,
$\r=P/Q$, $P$ and $Q$ relatively prime,
and we want to look for solutions such that $\f_x=\f_{x+Q}$.
Hence, we shall look for solutions of \equ(1.8) with $L=L_i=iQ$, $\r_L=\r$
and
$$ \f_x=\sum_{n=-[Q/2]}^{[(Q-1)/2]} \hat\f_n
e^{i2 \pi \r n x} \; ,\qquad \hat\f_n=\hat\f_{-n}\in\RRR\; ,
\qquad \hat\f_0 = 0\; .\Eq(1.10) $$
Note that the condition on $L$ allows to rewrite in a trivial way the field
$\f_x$ of \equ(1.10) in the general form \equ(1.9), by putting $\hat\f_n'=0$
for all $n$ such that $(2n\p)/L \not= 2\p\r m, \forall m$, and by relabeling
the other Fourier coefficients.

The conditions \equ(1.8) can be easily expressed in terms of the variables
$\hat\f_n$; if we define $\hat\r_n$ so that
$$\r_x(\f,\m)= \sum_{n=-[Q/2]}^{[(Q-1)/2]} \hat\r_n(\f,\m) e^{i2n\p \r x }
\; ,\Eq(1.11)$$
we get
$$\hat\f_n = \l\hat\r_n(\f,\m)\; ,\qquad n\not=0\; ,\quad n=-[Q/2],\ldots,
[(Q-1)/2] \; , \Eq(1.12)$$
$$\hat\r_0(\f,\m) =\r_L\; .\Eq(1.13)$$

Also the minimum condition \equ(1.7) can be expressed in terms of the
Fourier coefficients; we get that the $L\times L$ matrix
$$\bar M_{nm}\ \= \d_{nm} - \l {\dpr\over \dpr\hat\f'_n} \hat\r'_m(\f,\m)
\Eq(1.14)$$
has to be positive definite, if the field $\f$ satisfies \equ(1.12) and
\equ(1.13) and $\hat\r'_m(\f,\m)$ is defined analogously to $\hat\f'_m$
in \equ(1.9).
Hence, if we restrict the space of phonon fields to those of the
form \equ(1.10), we have to show that the $Q\times Q$ matrix
$$\tilde M_{nm}\ \= \d_{nm} - \l {\dpr\over \dpr\hat\f_n} \hat\r_m(\f,\m)
\Eq(1.15)$$
has to be positive definite, if the field $\f$ satisfies \equ(1.12) and
\equ(1.13).

\*

\sub(1.6) {\cs Remark.}
It is easy to show (by using the expansion described in \sec(3), for example)
that $\bar M_{nm}$ can be different from zero only if $2\p(n-m)/L$ is of the
form $2\p\r k$ for some $k$. However, we are not able to get good bounds on
all matrix elements $\bar M_{nm}$ not belonging (up to a relabeling of
indices) also to the matrix \equ(1.15);
therefore, in studying the minimum condition, we restrict ourselves to the
fields of the form \equ(1.10).

\*

\sub(1.7) {\cs Theorem.}
{\it Let $\r=P/Q$, with $P,Q$ relative prime integers, $L=L_i\=iQ$.
Then, for any positive integer $N$,
there exist positive constants $\e$, $\tilde\e$, $c$ and $K$,
independent of $i$, $\r$ and $N$, such that, if
$$ 0\le {4\p v_0\over \log(\tilde\e v_0\, L)} \le \l^2 \le \e\,{v_0^2
(1+\log v_0^{-1})^{-1}
\over K^N N! \log(c\,Q/v_0^4)}\; ,\Eq(1.16)$$
where
$$v_0=\sin (\p\r)\;,\Eq(3.25)$$
there exist two solutions $\f^{(\pm)}$ of \equ(1.8),
with $L=L_i$, $1-\m =\cos (\p\r)$ and $\r_L=\r$, of the form \equ(1.10).
The matrices $\tilde M$ corresponding to these solutions, defined as in
\equ(1.15), are positive definite.

Moreover, the Fourier coefficients $\hat\f_n^{(\pm)}$ verify, for $|n|>1$,
the bound
$$|\hat\f_n^{(\pm)}|\le \left({\l^2\over v_0 |n|}\right)^N |\hat\f_1^{(\pm)}
|\;.\Eq(1.17)$$

Finally, $\l\hat\f_1^{(\pm)}$ is of the form
$$\l\hat\f_1^{(\pm)}=\pm v_0^2 \exp \Big\{-{2\p v_0 + \b^{(\pm)}(\l,L)\over
\l^2}\Big\}\;,\Eq(1.18)$$
with
$$|\b^{(\pm)}(\l,L)| \le C\l^2\left(1+\log{1\over v_0}
\right)\;,\Eq(1.19)$$
where $C$ is a suitable constant.

The one-particle Hamiltonian ${\bf h}$ corresponding to this solution has a
gap of order $|\l \hat\f_1|$ around $\m$, uniformly on $i$.}

\*

\sub(1.8) The above theorem
proves that there are two stationary points of the ground state
energy in correspondence of a periodic function with period equal to
the inverse of the density, if the coupling is small enough and
the density is rational, and that these stationary points are local
minima at
least in the space of periodic functions with that period.
The energies associated to such minima are
different so that the ground state energy is not degenerate.

The theorem is proved by writing $\r_x(\f,\m)$
as an expansion convergent for small $\l$
and solving the set of equations \equ(1.12) by a contraction method.
As a byproduct we prove that the $\hat\f_n$ are
fast decaying, (see \equ(1.17)), so that $\f_x$
is really well approximated by its first harmonics (this remark
is important as the number of harmonics could be very large).

The results are uniform in the volume,
so they are interesting from a physical
point of view (a solution defined only for $|\l|\le O(1/L)$
should be outside any reasonable physical value for $\l$).
The case in [KL] for the half filled case is contained in Theorem \secc(1.7),
but in [KL] it is also proved that the solution is a global minimum.
On the other hand this case is quite special
(see Remarks \secc(2.5) in \sec(2)).

Finally the lower bound in \equ(1.16) is a large volume condition:
this is not a technical condition as, if the number of Fermions is odd,
there is Peierls instability only for $L$ large enough.
The upper bound for $\l$ in \equ(1.16) requires $\l$
to decrease as $Q$ increases:
in particular irrational density are forbidden.
This requirement is due to the discreteness of the lattice and
to Umklapp phenomena. Note that the dependence of the maximum
$\l$ allowed on $Q$ is not very strong as it is a logarithmic one.

The case of irrational densities (possible only
in the infinite volume limit), excluded by our theorem,
is physically interesting, but the existence of Peierls instability
in this case is proven only for large $\l$, [AL,BM].
In [BGM] $\r_x(\f,\m)$ is shown to be well defined for small $\l$
not only in the rational density case,
(in which the proof is almost trivial),
but also in the irrational case: in fact the small divisor
problem due to the irrationality of the density can be controlled
thanks to a Diophantine condition. However to solve the set of equations
\equ(1.12) we use a contraction method which is not trivially adaptable in
the latter case (see Remarks \secc(2.5) in \sec(2)).
The same kind of problem arises in proving the positive definiteness
of $\bar M_{nm}$ in the rational case (and this is the reason why
we are able to prove that the stationary points are local minima only
in the space of periodic functions with prefixed period).
As we said above, we do not know if such problems are
only technical or there is some physical reason for this to happen.

\pagina

\section(2,Solution of the self-consistence equation)

\sub(2.1) Let $\r=P/Q$, with $P,Q$ relatively prime integers such that
$0<P<Q$, and $L=L_i\=iQ$;
we have to look for a solution of \equ(1.12) and \equ(1.13), which is
well defined for $|\l|\le\e_0$, with $\e_0$ {\it independent} of $i$
(otherwise our solution is meaningless from a physical point of view).
As discussed in \sec(1.4), this means that
our solution has to be looked for in a class of functions for which
the one-particle Hamiltonian $\bf h$ has a gap around $\m$ of
width independent of $L$.
This class of functions is described by the following lemma, to be proved in
\sec(5.4).

\*

\sub(2.2) {\cs Lemma.} {\it Let $\f_x$ be a field of the form \equ(1.10),
$L=L_i$, $1-\m=\cos(\p\r)$, $|\l\hat \f_1|>0$ and $|\l\hat \f_n|\le
a|\l\hat \f_1|/|n|^N$ for some positive constants $a$ and $N$.
Then there exists $\e_0>0$, independent of $i$ and $\r$, such that,
if $|\l\hat\f_1|\le\e_0v_0^4/Q$, with $v_0=\sin(\p\r)$, the one-particle
Hamiltonian $\bf h$ has a gap of width $\ge |\l\hf1|/2$ around $\m$.
Moreover, $\hat\r_n(\f,\m)$ is a continuous function of $\l$, which\
converges to a continuous function of
$\l$ as $i\to\io$, and $\hat\r_0(\f,\m)=\r$.}

\*

\sub(2.3) We can write the self-consistence equation \equ(1.12) as
$$ \hf{n} = - \l^2 c_n(\s) \hf{n} + \l \tilde \r_n(\s,\F) \; , \quad
\s\=\l\hf1\; ,\quad \F\=\{\l\hf{n}\}_{|n|>1}\;,\Eq(2.1) $$
where $c_n(\s)$ depends on $\f$ only through
$\s$. We write $\hat\r_n$ as a perturbative expansion
in $\l$ (different from the power expansion in $\l$); this expansion is
described in \sec(3). If $|n|>1$, we are here defining $-\l c_n(\s) \hat\f_n$
the contribution to $\hat\r_n$ proportional to $\hf{n}$ of order $1$
in the expansion, while $-\s c_1(\s)$ is the contribution to $\hat\r_1$
proportional to $\s$ of order $\le 1$ in the expansion
(explicit expressions for $c_n(\s)$ and $c_1(\s)$ will
be given in \equ(4.9) and \equ(4.32) respectively);
$\tilde \r_n$ takes into account all the remaining terms of first
order plus all terms of order higher than 1.

The equation \equ(2.1) has of course the trivial solution $\hf{n}=0,
\forall n$,
but it is easy to see that this is not a local minimum, by using the expansion
for $\hat\r_n$ of \sec(3). Therefore we shall look for solutions such that
$\s\neq 0$, so that we can rewrite \equ(2.1) as
$$ \eqalignno{
& (1+\l^2 c_1(\s)) = {\l^2\tilde\r_1(\s,\F) \over \s } \; ,&\eq(2.2)\cr
& \F_n\= \l \hf{n} = {\l^2 \tilde\r_n(\s,\F) \over (1+\l^2 c_n(\s))} \; ,
\qquad |n|> 1 \; .&\eq(2.3)\cr}$$
Note that the equation for $n=-1$ does not appear simply because
$\r_{-1}=\r_{1}$, as a consequence of the condition
$\hat\f_n=\hat\f_{-n}\in\RRR$, see \equ(1.10).

\*

\sub(2.4) We prove Theorem \secc(1.7) in three steps as follows.\\
$\bullet$ We first study the self-consistence equation \equ(2.3),
considering $\s$ as a variable belonging to the interval
$$J = (\;-\exp (-\p\, v_0/\l^2)\;,\; \exp (-\p\, v_0/\l^2)\; )\;.\Eq(2.4)$$
We find a solution, that we denote $\F(\s)$, if $\l$ is small enough.\\
$\bullet$ We then prove that, if $L$ is large enough, the equation (in $\l$)
$$ 1+\l^2 c_1(\s) = {\l^2 \tilde \r_1 (\s,\F(\s))
\over \s} \Eq(2.5) $$
has two solutions $\s^{(\pm)}\in J$, of the form \equ(1.18).
Therefore $(\s^{(\pm)}(\l)/\l,$ $\F(\s^{(\pm)}(\l))/$ $\l)$ turn out to be
solutions of \equ(1.12), which verify, thanks to Lemma \secc(2.2),
\equ(1.13) with $L=L_i$.\\
$\bullet$ We finally prove that the Hessian matrices \equ(1.15)
corresponding to these two solutions are positive definite.

\*

\sub(2.5) {\cs Remarks.} The coefficient $\hf{1}$
has a privileged role with respect to the other coefficients.
In fact, as we shall see in \sec(5),
the properties of the system when only $\hf{1}$ is different from $0$
are very close to the properties of the case in which
all the coefficients are non vanishing.
This suggests that the ``important" equation is \equ(2.2),
so explaining the strategy outlined above.

The previous remark also implies that $1+\l^2 c_1(\s) \simeq 0$. It follows
that $1+\l^2 c_n(\s) \simeq 0$, for all $n$ such that $2\p\r n \simeq 2\p\r$
($\mod 2\p$). Since $\min_{|n|>1} |2\p\r n - 2\p\r|=2\p/Q$, we can expect
that our bounds will not be uniform in $Q$. This is the reason why Theorem
\secc(1.7) can not be extended to irrational density; at most one can hope
that a Diophantine condition on $\r$ is needed, but we have only been able
to prove that the $Q$ dependence can be substituted with a dependence on the
Diophantine constants in some of the bounds described below.

Note also that, if $Q=2$, the only equation to discuss
is just the equation \equ(2.2) with $\F=0$ and the r.h.s. equal to zero;
its solution is well known in this case, see [KL,LM] for example.
If $Q=3$, again \equ(2.2) is the only equation to discuss, but the r.h.s.
is different from zero; however it is easy to prove that the solution has
essentially the same properties as in the case $Q=2$.
Hence, in the following we shall consider only the case $Q\ge 4$.
The following lemma, furnishing a bound on the
constants $c_n(\s)$ and their derivatives, is proven in \sec(4.8).

\*

\sub(2.6) {\cs Lemma.} {\it There exists a constant $C$, independent of
$i$ and $\r$, such that, if $|n|\ge 2$,
$$|c_n(\s)| \le {C\over v_0}
\left(1+\log{1\over v_0}\right) \log Q\; ,\Eq(2.6)$$
$$\left|{\dpr c_n(\s)\over \dpr\s}\right|
\le {C\over v_0 |\s|}\; ,\Eq(2.6a)$$
}

\*

\sub(2.8)
Fixed $L=L_i$, $\F$ is a finite sequence of $Q-3$ elements, which can be
thought as a vector in $\RRR^{Q-3}$, which is a function of $\s$.
In order to study the equation \equ(2.2) for $\s$, we shall need a bound on
$\F$ and on the derivative of $\F$ with respect to $\s$. Hence we consider
the space $\FF=\CC^1(\,J\;,\RRR^{Q-3})$ of $C^1$-functions of $\s\in J$
with values
in $\RRR^{Q-3}$; the solutions of \equ(2.3) can be seen
as fixed points of the operator ${\bf T}_{\l}: \FF\to \FF$,
defined by the equation:
$$[{\bf T}_{\l}(\F)]_n(\s) =
{\l^2 \tilde\r_n(\s,\F(\s)) \over (1+\l^2 c_n(\s))} \; ,\Eq(2.7)$$

We shall define, for each positive integer $N$, a norm in $\FF$ in
the following way:
$$\|\F\|_\FF\= \sup_{|n|>1,\s\in J} \left\{
|n|^N \left[ |\s|^{-1} |\F_n(\s)|
+\left|{\dpr\F_n\over \dpr\s}(\s)\right|\right] \right\} \;.\Eq(2.8)$$

We shall also define
$$\BB = \{\F\in\FF : \|\F\|_\FF\le 1\}\; ;\Eq(2.9)$$
$$R(\F)_n(\s) = \tilde\r_n(\s,\F(\s))\;,\quad |n|\ge 2\;.\Eq(2.9a)$$

The following two lemmata, to be proved in \sec(5.5) and \sec(5.6),
respectively,
resume the main properties of $R(\F)$.

\*

\sub(2.9) {\cs Lemma.} {\it There are two constants $C_1>1$ and $C_2$,
independent of $i$, $\r$ and $N$, such that, if $\F,\F'\in\BB$ and
$$C_1 Q v_0^{-4} |\s|[1+\log (v_0^2/|\s|)]\le 1\;,\Eq(2.9b)$$
then
$$\|R(\F)-R(\F')\|_\FF \le {C_2 3^N N! \over v_0}
\left(1+\log{1\over v_0}\right)
\|\F-\F'\|_\FF\; .\Eq(2.10)$$}

\*

\sub(2.10) {\cs Lemma.} {\it There is $C>1$, such that, if
$$C Q v_0^{-3} |\s|^{1/2} [1+\log (v_0^2/|\s|)] \le 1\;,\Eq(2.9c)$$
then
$$\|R(0)\|_\FF \le {C\over v_0} \left(1+\log{1\over v_0}\right)
\sup_{|n|>1} \left\{ |n|^N \Big({|\s|\over v_0^2}\Big)^{|n|\over 10}
\right\} \; .\Eq(2.11)$$
}

\*

\sub(2.11) {\cs Lemma.} {\it There are $\e,c,K$, independent of $i$, $\r$
and $N$, such that, if $\s\in J$ and
$$\l^2 \le \e\,{v_0^2 (1+\log v_0^{-1})^{-1}
\over K^N N! \log(c\,Q/v_0^4)}\;,\Eq(2.11a)$$
there exists a unique solution $\F\in\BB$ of \equ(2.3);
moreover the solution satisfies the bound
$$\|\F\|_\FF \le \left({\l^2\over v_0}\right)^N\; .\Eq(2.12)$$
}

\*

\sub(2.12) {\it Proof of Lemma {\secc(2.11)}.}
It is easy to see that, if $\s\in J$, the
conditions on $\s$ of Lemma \secc(2.9) and Lemma \secc(2.10) are
satisfied, if
$$\l^2 \le \e_0/\log(c Q/v_0^4)\; ,\Eq(2.13)$$
with suitable values of $\e_0$ and $c$. Moreover, if $\e_0\le \e_1 v_0
(1+\log v_0^{-1})^{-1}$ and $\e_1$ is chosen small
enough, \equ(2.6) and \equ(2.13) imply that $\l^2|c_n(\s)|\le 1/2$,
so that, by using \equ(2.6a), \equ(2.7) and Lemma \secc(2.9), we have that,
if $\F\in\BB$,
$$ \|{\bf T}_{\l} (\F)\|_\FF \le 4 \l^{2} \left(1+\l^2{C\over v_0}\right)
\left[ \|R(0)\|_\FF + {C_2\over v_0}\left(1+\log{1\over v_0}\right)
3^N N! \|\F\|_\FF\right]\; . \Eq(2.14)$$

Therefore, by \equ(2.11) and \equ(2.4), there exist constants $C_3$ and $C_4$,
such that, if $\e_1\le \e v_0 (C_4^N N!)^{-1}$ and $\e$ is small enough,
$$ \|{\bf T}_{\l} (\F)\|_\FF
\le {C_3\l^2\over v_0^2} \left(1+\log{1\over v_0}\right)
\left[3^N N! + \sup_{|n|>1}
|n|^N \exp \Big(-{\p v_0 |n|\over 10 \l^2 }\Big)
\right] \le 1\; . \Eq(2.15)$$

Moreover, by \equ(2.10), if $\F,\F'\in\BB$ and
similar conditions on $\l$ are satisfied, we have
$$\|{\bf T}_{\l}(\F)-{\bf T}_{\l}(\F')\|_\FF \le {C_5^N N!\l^2\over v_0^2}
\left(1+\log{1\over v_0}\right) \|\F-\F'\|_\FF \le
\fra12 \|\F-\F'\|_\FF \; .\Eq(2.16)$$

The bounds \equ(2.15) and \equ(2.16) imply that $\BB$ is invariant
under the action of ${\bf T}_{\l}$ and that ${\bf T}_{\l}$ is a
contraction on $\BB$. Hence, by the contraction mapping principle, there is
a unique fixed point $\bar\F$ of ${\bf T}_{\l}$ in $\BB$, which can
be obtained as the limit of the sequence $\F^{(k)}$ defined through the
recurrence equation $\F^{(k+1)}={\bf T}_{\l}(\F^{(k)})$, with any initial
condition $\F^{(0)}\in\BB$. If we choose $\F^{(0)}=0$, we get, by
\equ(2.16),
$$\|\bar\F\|_\FF \le \sum_{i=1}^\io \|\F^{(i)}-\F^{(i-1)}\|_\FF
\le \sum_{i=1}^\io {1\over 2^{i-1}} \|\F^{(1)}\|_\FF \le
\|\F^{(1)}\|_\FF \; .\Eq(2.17)$$
On the other hand, by \equ(2.11),
$$\|\F^{(1)}\|_\FF = \|{\bf T}_{\l}(0)\|_\FF \le
{C_6^N N! \l^2\over v_0^2} \left(1+\log{1\over v_0}\right)
\left({\l^2\over v_0}\right)^N\; ,\Eq(2.18)$$
which immediately implies the bound \equ(2.12), if $\e_1\le \e
v_0(C_6^N N!)^{-1}$, with $\e$ small enough. \qed

\*
\sub(2.13) Let us now consider the equation \equ(2.5). We want to prove that
it has two solutions of the form \equ(1.18), if $\s\in J$ and $L_i$ is large
enough.
In order to achieve this result, we need some detailed properties of the
function $c_1(\s)$, which are described in the following Lemma \secc(2.13a),
to be proved in \sec(4.9). We need also the bounds on $\tilde\r_1(\s,\F(\s))$
and its derivative with respect to $\s$, contained in Lemma \secc(2.13b), to
be proved in \sec(5.7).

\*
\sub(2.13a) {\cs Lemma.}
{\it There is a constant $C$, such that, if
$${v_0\over L_i |\s|} \le \tilde\e \le {1\over 8\p}
\; ,\qquad {|\s|\over v_0^2}\le 1\;,\Eq(2.18a)$$
then
$$-c_1(\s) = {1\over 2\p v_0} \left[ \log {v_0^2\over |\s|} + r_1(\s)
\right]\;,\Eq(2.19)$$
with
$$\eqalign{
|r_1(\s)| &\le C \left(1+\log{1\over v_0}\right)\;,\cr
\left|{\dpr r_1(\s)\over \dpr\s}\right| &\le C \left( {1\over v_0^2}
+{\tilde\e\over |\s|} \right)\;.\cr}\Eq(2.19a)$$
}

\*

\sub(2.13b) {\cs Lemma.}
{\it If $\s\in J$, $\l$ satisfies the inequality \equ(2.11a), with
$\e$ small enough, $\F(\s)$ is
the solution of the equation \equ(2.3) described in Lemma \secc(2.11) and
$$r_2(\s)\= {2\p v_0 \tilde \r_1 (\s,\F(\s))\over \s}\;,\Eq(2.19b)$$
then there is a constant $C$, such that
$$\eqalign{
|r_2(\s)| &\le C 
\left[\left({|\s|\over
v_0^2}\right)^{1/4} + \left({\l^2\over v_0}\right)^N \right]
\;,\cr
\left|{\dpr r_2(\s)\over \dpr\s}\right| &\le {C
\over |\s|}
\left[\left({|\s|\over
v_0^2}\right)^{1/4} + \left({\l^2\over v_0}\right)^N \right]
\;.\cr}\Eq(2.19c)$$
}

\*
\sub(2.14) {\cs Lemma.}
{\it There exist positive constants $\e$, $\tilde\e$, $c$ and $K$,
independent of $i$, $\r$ and $N$, such that,
if $\l$ satisfies the inequalities \equ(1.16),
there are two solutions $\s^{(\pm)}(\l)\in J$ of equation \equ(2.5)
of the form \equ(1.18).}

\*

\sub(2.15) {\it Proof of Lemma {\secc(2.14)}.}
By using the definitions of $r_1(\s)$ and $r_2(\s)$ given in \equ(2.19) and
\equ(2.19b), we can write the equation \equ(2.5) in the form
$$F(\s)\=\log {v_0^2\over |\s|} -{2\p v_0\over \l^2}+r(\s)
=0\;,\Eq(2.20)$$
where $r(\s)=r_1(\s)+r_2(\s)$.

Let us now suppose that $\l$ satisfies the
inequalities \equ(2.11a) and that $\s$ belongs to the interval
$$\tilde J= \left( v_0^2 e^{-4\p v_0/\l^2}\;,\; v_0^2 e^{-\p
v_0/\l^2} \right)\subset J\;.\Eq(2.21)$$
If $L_i$ is large enough and the constant $\e$ in \equ(2.11a) is chosen small
enough, the conditions \equ(2.18a) of Lemma \secc(2.13a) are satisfied, for
$\s\in \tilde J$, and
$${4\p v_0\over \l^2} \le \log(\tilde\e v_0 L_i)\;.\Eq(2.21a)$$
Moreover, if $\tilde \e$ and $\e$ (hence $|\s|v_0^{-2}$) are small enough,
$$\left| {\dpr r(\s)\over \dpr\s} \right| \le {1\over 2}
\left| {\dpr \over \dpr\s} \log {v_0^2\over\s}\right|\;;\Eq(2.22)$$
hence $F(\s)$ is a monotone decreasing function of $\s$ in $\tilde J$.
If we define
$$\s^* = v_0^2 e^{-2\p v_0/\l^2}\;,\qquad M=\sup_{\s\in\tilde J} |r(\s)|\;,
\Eq(2.23)$$
we have that $F(\s^*\exp (-2M))>0$ and $F(\s^*\exp (2M))<0$. Moreover, the
interval $(\s^*\exp (-2M)),\s^*\exp (2M)))$ is contained in $\tilde J$, if
$\e$ is small enough, since the bounds \equ(2.19a) and \equ(2.19c)
imply that $M\le C (1+\log v_0^{-1})$.
Hence there is a unique solution $\s^{(+)}(\l)$ of
\equ(2.20) in $\tilde J$, which can be written as
$$\s^{(+)}(\l)= v_0^2 e^{-{2\p v_0+\b^{(+)}(\l)\over \l^2} }\;,\Eq(2.24)$$
with $|\b^{(+)}(\l)| \le C\l^2(1+\log v_0^{-1})$.

In the same manner, we can show that there is solution $\s^{(-)}(\l)$
in the interval
$$ (-v_0^2 e^{-{\p v_0 \over \l^2}}\;,\;
-v_0^2 e^{-{4\p v_0\over \l^2}}) \subset J \; , \Eq(2.25) $$
with the same properties. \qed

\*

\sub(2.16) {\cs Lemma.} {\it The constants $\e$, $\tilde\e$, $c$ and $K$,
appearing in \equ(1.16), can be chosen so that the Hessian matrix \equ(1.15)
is positive definite.}

\*

\sub(2.17) The proof of Lemma \secc(2.16) is in \sec(5.8).
This completes the proof of Theorem \secc(1.7).

\pagina

\section(3,Graph formalism)

\sub(3.1) In this section we shall describe the expansion of $\r_x(\f,\m)$,
used to get the results of this paper.

Let us consider the operators
$\psi_{\xx}^{\pm}=e^{tH}\psi_x^{\pm}e^{-Ht}$, with $\xx=(x,t)$,
$-\b/2\le t \le \b/2$ for some $\b>0$; on $t$ antiperiodic boundary
conditions are imposed. As explained, for example, in [BGM], there is a simple
(well known) relation between $\r_x(\f,\m)$ and the {\sl two-point Schwinger
function}, defined by
$$S^{L,\b}(\xx;\yy) = {{\rm Tr} \left[\exp(-\b H)
\left(\theta(x_0>y_0)\psi^-_{\xx} \psi^+_{\yy}-
\theta(x_0<y_0)\psi^-_{\xx} \psi^+_{\yy}\right)\right]
\over {\rm Tr} \left[\exp(-\b H)\right]} \; ,\Eq(3.1)$$
given by
$$ \r_x = - \lim_{\b\to\io}\lim_{\t\to 0^-} {1\over L}
S^{L,\b}(x,\t;x,0)\; . \Eq(3.2)$$
In [BGM], which we shall refer to for more details, it is also explained that
the two-point Schwinger function can be written as
$$ S^{L,\b}(\xx;\yy) = \lim_{M\to\io} {\int P(d\psi)\,
e^{\VV(\psi)}\,\psi^-_{\xx}\psi^+_{\yy}
\over\int P(d\psi)\,e^{\VV (\psi)}} \; , \Eq(3.3) $$
where $\psi^\pm_{\xx}$ are now anticommuting Grassmanian variables and
$P(d\psi)$ is a {\sl Grassmanian Gaussian measure}, formally defined by
$$P(d\psi) = \Big\{ \prod_{\kk\in{\cal D}_{L,\b}} (L\b\hat g(\kk)) \Big\}
\exp \left\{-\sum_{\kk\in{\cal D}_{L,\b}} (L\b \hat g(\kk))^{-1} \psi_\kk^+
\psi_\kk^- \right\} d\psi^- d\psi^+ \; ,\Eq(3.4)$$
$\kk=(k,k_0)$, ${\cal D}_{L,\b}\={\cal D}_L \times {\cal D}_\b$,
${\cal D}_L\=\{k={2\pi n/L}, n\in \ZZZ, -[L/2]\le n \le [(L-1)/2]\}$,
${\cal D}_\b\=\{k_0=2(n+1/2)\pi/\b, n\in \ZZZ, -M\le n \le M-1\}$,
in the limit $M\to\io$,
$$ \hat g(\kk) = {1\over -ik_0+\cos p_F - \cos k }\Eq(3.5)$$
is the {\sl propagator} or the {\sl covariance of the measure},
$p_F=\p\r$ is the {\sl Fermi momentum}, defined so that $\cos p_F =1-\mu$, and
$$ \VV(\psi)=\sum_{x\in\L} \int_{-\b/2}^{\b/2} dx_0
\Big[\l \f_x\psi_\xx^+ \psi^-_\xx \Big] \; .\Eq(3.6) $$

If we insert \equ(1.10) in the r.h.s. of \equ(3.6), we get
$$\VV(\psi) =
\sum_{n=-[Q/2]}^{[(Q-1)/2]} {1\over L\b} \sum_{\kk\in {\cal D}_{L,\b}}
\, \l \hat \f_n \, \psi^+_\kk \psi^-_{\kk+2n\pp_F} \; ,\Eq(3.7)$$
where $\pp_F=(p_F,0)$ and $k+2np_F$ is of course defined modulo $2\p$.

\*

\sub(3.2) Note that $\hat g(\kk)^{-1}$ is small for $\kk\simeq \pm\pp_F$.
Hence there is no hope to
treat perturbatively the terms with $n=\pm 1$ and $\kk$ near $\mp \pp_F$,
but we can at most expect that the interacting measure is a perturbation of
the measure (whose covariance is not singular at $\kk=\pm\pp_F$)
$$ \eqalign{
\bar P_\l(d\psi) \= {1\over \NN} P & (d\psi) \cr &
\exp \left\{ \l\hat \f_1 {1\over\b}\sum_{k_0\in {\cal D}_\b}
\fra1L \sum_{k\in I_-} \Big[\psi^+_\kk \psi^-_{\kk+2\pp_F} +
\psi^+_{\kk+2\pp_F} \psi^-_\kk\Big] \right\}\; , \cr} \Eq(3.8)$$
where $\NN$ is a normalization constant and $I_-$ is a small interval
centered in $-p_F$, so small that $I_-\cap I_+ =
\emptyset$, if $I_+ \= \{k=\bar k+2p_F, \bar k\in I_-\}$.

This remark suggests to apply a multiscale expansion to the integral
\equ(3.3), in order to treat in a different way the momenta near $\pm p_F$
and the others. This procedure was applied in [BGM] to study systems of
electrons in presence of a potential of the form $\bar\f(2px)$, with $\bar\f$
$2\p$-periodic, $p/\p$ an irrational Diophantine number and $p_F=mp$, $m$
arbitrary integer.
In [BGM] the aim was mainly to get the best possible results about
the dependence of the two-point Schwinger function on $\l$ and we found it
useful to realize the multiscale expansion by dividing the momenta near $p_F$
into a number of ``slices'' of order $|\log\l|$.

This expansion could be applied also the case $p_F=p$ with $\r=p_F/\p$
rational, without no important difference, and we could get immediately
Lemma \secc(2.2). However, in this paper we prefer to use a simpler expansion
into only two scales; this expansion gives weaker results about the dependence
on $\l$, but it is sufficient in order to prove Lemma \secc(2.2) and it is
more suitable for studying the equation \equ(1.12).

\*

\sub(3.3) Let us introduce a smooth positive function $f_0(k')$ on the
one dimensional torus $\ttt^1$, such that
$$ f_0(k') = \cases{
1 \; , \qquad & if $\qquad \|k'\|_{\ttt^1} \le t_0/2 \; , $ \cr
0 \; , \qquad & if $\qquad \|k'\|_{\ttt^1} \ge t_0   \; , $ \cr}
\Eq(3.9) $$
where
$$t_0=\min\{p_F/2,(\p-p_F)/2\}\;. \Eq(3.10)$$
and the norm $\|k'\|_{\ttt^1}$ on $\ttt^1$ is defined so that
$\|k'\|_{\ttt^1}=|k'|$, if $k'\in[-\p,\p)$.
Then we write
$$\hat f_1(k)=1-f_0(k+p_F)-f_0(k-p_F) \;,\qquad \hat f_0(k)=
1-\hat f_1(k) \;, \Eq(3.11) $$
so that \equ(3.5) becomes
$$ \hat g(\kk) = \hat g^{(1)} (\kk) + \hat g^{(0)}(\kk) = \sum_{h=0,1}
{\hat f_h(k) \over -ik_0+\cos p_F - \cos k } \; , \Eq(3.12) $$
and, for $h=0$, we define
$$ \hat g^{(0)}(\kk) = \sum_{\o=\pm 1}
\hat g^{(0)}_{\o}(\kk) \; , \Eq(3.13) $$
where, if $\kk'=\kk-\o \pp_F$,
$$\hat g^{(0)}_{\o}(\kk'+\o \pp_F) \= \tilde g^{(0)}_{\o}(\kk') =
{f_0(k') \over -ik_0+\cos p_F - \cos (k'+\o p_F) } \; , \Eq(3.14) $$
with $\kk'=(k',k_0)$;
we set also $\tilde g^{(1)}_{1,1}(\kk')\=\hat g^{(1)}(\kk)$,
with $k=k'+p_F$, and $f_1(k')=\hat f_1(k)$, in order to simplify the
notations in the following sections.

\*

\sub(3.4) We can associate with the decomposition \equ(3.12) of $\hat g(\kk)$
a decomposition of the Grassmanian Gaussian measure $P(d\psi)$ into
a product of two independent Grassmanian Gaussian measures:
$$ P(d\psi)= P(d\psi^{(1)}) P(d\psi^{(0)}) \; ,\Eq(3.15)$$
if $P(d\psi^{(i)})$ is defined as in \equ(3.4), with $\hat g^{(i)} (\kk)$
in place of $\hat g(\kk)$.

If we insert \equ(3.15) in \equ(3.3) and we perform the integration over
the field $\psi^{(1)}$, it is easy to show (see [BGM], \S4) that
$$S^{L,\b}(\xx;\yy) = g^{(1)}(\xx;\yy)+
K^{(0)}_{\phi,\phi}(\xx;\yy) + S^{(0)}(\xx;\yy) \; ,\Eq(3.16)$$
where $g^{(1)}(\xx;\yy)=(L\b)^{-1} \sum_{\kk\in {\cal D}_{L,\b}}
\hat g^{(1)}(\kk)\,\exp[-i\kk\cdot(\xx-\yy)]$ and
$$\eqalign{
S^{(0)}(\xx;\yy) &= {\partial^2 \over \dpr \phi_\xx^+\partial\phi^-_\yy}
{1\over\NN_0}\,\int P(d\psi^{(0)})\cr
& e^{\int d\xx \big( \phi^+_\xx\psi^{(0)-}_\xx  +
\psi^{(0)+}_\xx \phi^-_\xx \big)}  e^{\VV^{(0)}(\psi^{(0)})
+ W^{(0)}(\psi^{(0)},\phi) } \; \Big|_{\phi^+=\phi^-=0}
\; ;\cr}\Eq(3.17)$$
in \equ(3.17) $\int d\xx$ is a shortcut for $\sum_{x\in\L} \int_{-\b/2}^{\b/2}
dx_0$,\hfill\break
$\NN_0 = \int P(d\psi^{(0)})\,\exp [\VV^{(0)}(\psi^{(0)})]$,
$\{\phi_\xx^\pm\}$ are Grassmanian variables anticommuting with
$\{\psi_\xx^\pm\}$ and $\VV^{(0)}(\psi^{(0)})$, the {\sl effective potential
on the small momenta scale}, can be easily represented as a series in $\l$, as
well as the function $W^{(0)}(\psi^{(0)},\phi)$ and the function
$K^{(0)}_{\phi,\phi}(\xx;\yy)$ appearing in \equ(3.16).

A precise description of these series in terms of Feynman graphs can be
found in [BGM]; see in particular the equations (4.3), (4.4)
and (4.6) in [BGM].
Here we want only to stress that the involved graphs are chains
formed by vertices, associated with the Fourier components of the
potential $\hat\f_n$, connected through lines, associated to the
propagators $g^{(1)}$; hence
by using the bound \equ(3.28) below, it is easy to prove that these series are
convergent, uniformly in $L$ and $\b$. On the contrary, the series obtained
by integrating the field $\psi^{(0)}$ do not have this property, for the
reason explained before, and we have to look for a different expansion, based
on the idea outlined in \sec(3.2).

We can associate with the decomposition \equ(3.13) of $\hat g^{(0)}(\kk)$
a decomposition $\psi^{(0)}_\kk = \psi^{(0)}_{\kk,+} + \psi^{(0)}_{\kk,-}$
of the field $\psi^{(0)}_\kk$. The support properties of $f_0(k')$,
see \equ(3.9), and the definition \equ(3.14) imply that the field
$\psi^{(0)}_{\kk,\o}$ has support on the set
$\{\kk=\kk'+\o\pp_F:f_0(k')\not=0\}$ and that the supports
of $\psi^{(0)}_{\kk,+}$ and $\psi^{(0)}_{\kk,-}$ are disjoint.

The idea is to modify the {\sl free measure} $P(d\psi^{(0)})$ by multiplying
it by the terms present in $\VV^{(0)}(\psi^{(0)})$,
(see [BGM], \S3), which couple the variables
$\psi^{(0)}_{\kk,-}$ and $\psi^{(0)}_{\kk+2\pp_F,+}$; then we expand the
integral by using the new measure as the free measure.
The new graphs differ from the previous
ones for two respects; first of all they are not singular anymore at
$\kk=\pm\pp_F$, but they are bounded by $C|\l\hat\f_1|^{-1}$, see below;
moreover the two propagators exiting and entering
in the same vertex can not have both the momentum equal to $\pm\pp_F$.
As we shall see, these two properties are
sufficient to control the expansions.

The two properties of the new free measure described above are realized also
if we only extract from $\VV^{(0)}(\psi^{(0)})$ the first order terms
coupling $\psi^{(0)}_{\kk,-}$ and $\psi^{(0)}_{\kk+2\pp_F,+}$.
It is easy to see that these
terms are equal to $\l\hat\f_1 F_\s^{(0)}(\psi^{(0)})$, with
$$F_\s^{(0)}(\psi^{(0)}) = \sum_{\o=\pm1}{1\over L\b}
\sum_{\kk'\in {\cal D}_{L,\b}} \,
\psi^{(0)+}_{\kk'+\o\pp_F,\o}\psi^{(0)-}_{\kk'-\o\pp_F,-\o} \; .
\Eq(3.18)$$
Hence we define
$$\tilde P(d\psi^{(0)})= {1 \over \NN} P(d\psi^{(0)})
e^{\l\hat\f_1 F_\s^{(0)}(\psi^{(0)})}\; ,\Eq(3.19)$$
where $\NN$ is a suitable constant, and
$$\tilde\VV^{(0)}(\psi^{(0)}) = \VV^{(0)}(\psi^{(0)}) -
\l\hat\f_1 F_\s^{(0)}(\psi^{(0)})\; .\Eq(3.20)$$

By proceeding as in [BGM], \S3, one can show that
the Grassmanian integration $\tilde P(d\psi^{(0)})$ has propagator
$$ g^{(0)}(\xx;\yy) = \sum_{\o,\o'=\pm1}
e^{-i(\o x - \o' y)p_F}\,
g^{(0)}_{\o,\o'}(\xx;\yy) \; , \Eq(3.21)$$
with
$$ g^{(0)}_{\o,\o'}(\xx;\yy)={1\over L\b} \sum_{\kk'\in {\cal D}_{L,\b}} \,
e^{-i\kk'\cdot(\xx-\yy)} \tilde g_{\o,\o'}^{(0)}(\kk')\; , \Eq(3.22)$$
$$ \eqalign{
\tilde g_{\o,\o}^{(0)}(\kk') & = {[-ik_0-F_1(k')-\o F_2(k') ]
\, f_0(k')\over
[-ik_0 - F_1(k')]^2 - [ F_2^2(k') + \s^2_0(k')]} \; , \cr
\tilde g_{\o,-\o}^{(0)}(\kk') & = { [\s_0(k') ] \, f_0(k')\over
[-ik_0 - F_1(k')]^2 - [ F_2^2(k') + \s^2_0(k')]} \; , \cr} \Eq(3.23) $$
$\s_0(k')=\s f_0(k')$, if $\s=\l\hf1\neq0$, and (see \equ(3.25) for the
definition of $v_0$)
$$ F_1(k') = (\cos k'-1)\cos p_F \; , \qquad
F_2(k')=v_0\,\sin k' \;.\Eq(3.24) $$

\*

\sub(3.5) {\cs Remark.} Note that, if $|k'|\le t_0$, $2(1-\cos k')/|\sin k'| =
2|\tan (k'/2)| \le 2\tan (t_0/2) \le |\tan p_F|$; hence
$$ \left| F_1(k') \right| \le {1\over2} \left| F_2(k') \right| \;
, \qquad \hbox{for } |k'|\le t_0 \; .\Eq(3.26)$$
It immediately follows that
$$ \eqalign{
|\tilde g_{\o,\o}^{(0)}(\kk')| & \le C {\sqrt{k_0^2+(v_0k')^2}\over
k_0^2+(v_0k')^2+\s^2} f_0(k')\; ,\cr
|\tilde g_{\o,-\o}^{(0)}(\kk')| & \le C {|\s|\over k_0^2+(v_0k')^2+\s^2}
f_0(k')\; .\cr} \Eq(3.27)$$
where $C$ denotes a suitable constant. From now on, for simplifying the
notation, the symbol $C$ will be used everywhere to denote a generic constant,
that we do not need to better specify.

It is also easy to prove that
$$|\tilde g^{(1)}_{1,1}(\kk')| \le
C{1-f_0(k') \over \sqrt{|k_0|^2 + (v_0 k')^2}}\; ,\Eq(3.28)$$

\*

\sub(3.6) We now insert \equ(3.19) and \equ(3.20) in \equ(3.17) and represent
the result of the integration in terms of Feynman graphs, by using
$\tilde P(d\psi^{(0)})$ as the free measure and
$\tilde\VV^{(0)}(\psi^{(0)})$ as the effective potential;
then we apply \equ(3.2).
By proceeding as in [BGM], it is easy to show that we get an expansion for
$\r_x$, which can be described in the following way.

\*

\sub(3.7) A graph $\th$ of order $q\ge 1$ is a chain of $q+1$ lines
$\ell_1,\ldots,\ell_{q+1}$ connecting
a set of $q$ ordered points ({\sl vertices}) $v_1,\ldots,v_q$, so that
$\ell_i$ enters $v_i$ and $\ell_{i+1}$ exits from $v_i$,
$i\le q$; the lines $\ell_1$ and $\ell_{q+1}$ are the
{\sl external lines} of the graph and both have
a free extreme, while the others are the {\sl internal lines};
we shall denote ${\rm int}(\th)$ the set of all internal lines.
We say that $v_i<v_j$ if $v_i$ precedes $v_j$ and we denote
$v_j'$ the vertex immediately following $v_j$, if $j<q$.
We denote also by $\ell_v$ the line entering the vertex $v$,
so that $\ell_i \= \ell_{v_i}$, $1\le i \le q$.
We say that a line $\ell$ emerges from a vertex $v$ if
$\ell$ either enters $v$ ($\ell=\ell_{v}$) or exits from $v$
($\ell=\ell_{v'}$). By a slight abuse of notation,
if $v=v_q$, we still denote by $\ell_{v'_q}$ the line $\ell_{q+1}$
exiting from $v_q$ even if there is no vertex $v_{q+1}$.

We shall say that $\th$ is a {\sl labeled graph} of order $q\ge 1$,
if $\th$ is a graph of order $q$, to which the following {\sl labels} are
associated:\\
$\bullet$ a label $n_v$ for each vertex,\\
$\bullet$ a {\sl frequency} (or {\sl scale}) label $h_\ell$
for each (internal or external) line, with the constraint
that, if $n_v=\pm1$ for some $v$, then
$h_{\ell_{v}}=h_{\ell_{v'}}=0$ is not possible,\\
$\bullet$ for each line $\ell$, two labels $\o^1_\ell$, $\o^2_\ell$,
such that $\o^1_\ell$ $=$ $\o^2_\ell$ $=1$ if $h_{\ell}=1$,\\
$\bullet$ a momentum $k_{\ell_1}=k=k'+\o_1 p_F$ for the first line,\\
$\bullet$ a momentum
$$k_{\ell_v}=k' +\sum_{\tilde v < v} \left[ 2n_{\tilde v}p_F +
( \o^2_{\ell_{\tilde v}} - \o^1_{\ell_{\tilde v'}} )p_F \right]
\Eq(3.29)$$
for each internal line,\\
$\bullet$ a momentum
$$k_{\ell_{q+1}}=k' +\sum_{v\in\th} \left[ 2n_{\tilde v}p_F +
( \o^2_{\ell_{\tilde v}} - \o^1_{\ell_{\tilde v'}} )p_F \right]
\Eq(3.30)$$
for the last line.

If $\tilde g_{\o_{\ell}^1,\o_{\ell}^2}^{(h_\ell)}(\kk_{\ell}')$
denotes the propagator associated with the line $\ell$, we will
use the shorthand $\tilde g_{\ell}
=\tilde g_{\o_{\ell}^1,\o_{\ell}^2}^{(h_\ell)}(\kk_{\ell}')$.

Let us call $\TT_{n,q}$ the set of the labeled graphs of order $q$
and such that
$$ \sum_{v\in\th} 2n_v p_F + \sum_{\ell\in\th}
\left( \o^2_\ell-\o^1_\ell \right) p_F = 2np_F \quad \mod 2\p\; .\Eq(3.31)$$
Then, if $\hat\r_n(\f,\mu)$ is defined as in
\equ(1.11), we have
$$\eqalign{
\hat\r_n(\f,\mu) &= \lim_{\b\to\io} \left[
-{1\over L\b} \sum_{\kk'\in {\cal D}_{L,\b}} \sum_{\o=\pm 1}\d_{n,\o}
\tilde g_{-\o,\o}(\kk') \;+\;\sum_{q=1}^\io \r_n^q(\s,\F)\right] \;,\cr
\r_n^q(\s,\F)&=\sum_{\th\in\TT_{n,q}}\Val(\th) \; ,\cr}\Eq(3.32)$$
where
$$\Val(\th) = -{1\over L\b} \sum_{\kk'\in {\cal D}_{L,\b}}
\Big( \prod_{i=1}^{q+1} \tg{\ell_i} \Big)
\Big( \prod_{v\in\th} \l \hf{n_v} \Big) \; . \Eq(3.33)$$
Hence, the function $\tilde\r_n(\s,\F)$ defined in \equ(2.1) can be written
as
$$\eqalign{
\tilde\r_n(\s,\F) &= \lim_{\b\to\io} \sum_{q=1}^\io \tilde\r_n^q(\s,\F)\cr
\tilde\r_n^q(\s,\F) &= \r_n^q(\s,\F)\; ,\quad \hbox{if\ }q\ge 2\cr
\tilde\r_n^1(\s,\F) &= \sum_{\th\in\TT_{n,1}} (1-\d_{n,n_v})
\Val(\th)\; .\cr}\Eq(3.34)$$
after substituting in the r.h.s. of \equ(3.33) $\l \hf{n_v}$ either with
$\F_{n_v}$, if $|n_v|>1$, or with $\s$, if $|n_v|=1$.

\pagina

\section(4,First order graphs)

\sub(4.1) In this section we study the first order contributions to the
density, \ie the terms corresponding to graphs with only one vertex in the
perturbative expansion \equ(3.32), calculated in the limit $\b\to\io$.
For these graphs we have, if $L=L_i=iQ$,
$$\eqalign{
&\lim_{\b\to\io}\Val(\th) =\cr
&\quad - \l \hf{m} {1\over L}\sum_{k'\in \DD'_L}
\int_{-\io}^{\io} {dk_0\over2\p}
\, \tilde g_{\o_1,\o_1'}^{(h)}(\kk')\,
\tilde g_{\o_2,\o_2'}^{(h')}(\kk'+(2m+\o_1'-\o_2)\pp_F) \; ,\cr} \Eq(4.1) $$
where $\DD'_L$ is the set of possible values of the variable $k'$ introduced
before \equ(3.14) as the difference between the ``space momentum'' $k$ and
$\pm p_F$. Since $p_F=\p\r=\p P/Q=(2\p/L)(iP/2)$, we have
$$\DD'_L = \{ k'={2\p\over L} (n+\d/2), n\in \ZZZ, -[L/2]\le n \le [(L-1)/2]
\}\;,\Eq(4.1a)$$
where $\d=1$, if $iP$ (the number of particles) is odd, while $\d=0$,
if $iP$ is even.

Note that the value of $\d$ will be in general not relevant, except in the
proof of Lemma \secc(2.13a) in \sec(4.9), the only place were there is a non
trivial dependence on the volume.

Note also that, if the graph value \equ(4.1) contributes to
$\hat\r_n(\f,\m)$, then
$$2mp_F = 2np_F + (\o_1-\o_1'+\o_2-\o_2')p_F\;\mod 2\p\;.\Eq(4.2)$$

The r.h.s of \equ(4.1) can be easily bounded, by
using \equ(3.28) and \equ(3.27) and the remark that $\lim_{p_F\to 0} t_0/v_0
=1/2$. If $h=h'=1$, one gets, for any integer $r$,
$$\eqalign{
&{1\over L}\sum_{k'\in \DD'_L}\int_{-\io}^{\io} {dk_0\over2\p}
\left| \tilde g_{1,1}^{(1)}(\kk') \,
\tilde g^{(1)}_{1,1}(\kk'+2r\pp_F) \right| \le\cr
&C\int_{-\p}^{\p} dk' \int_{-\io}^{\io} dk_0
{[1-f_0(k')]\over \sqrt{k_0^2+(v_0 k')^2}}
{[1-f_0(k'+2rp_F)]\over \sqrt{k_0^2+v_0^2(k'+2rp_F)^2}} \le\cr
&C\int_{t_0}^{2\p} dk' \int_0^{\io} {dk_0\over k_0^2+(v_0 k')^2}
\le {C\over v_0}\left(1+\log {1\over v_0}\right)\; .\cr}\Eq(4.3)$$
If $h=0$, $h'=1$, for any $\o,\o'$, one gets
$$\eqalign{
&{1\over L}\sum_{k'\in \DD'_L}\int_{-\io}^{\io} {dk_0\over2\p}
\left| \tilde g_{\o,\o'}^{(0)}(\kk') \,
\tilde g^{(1)}_{1,1}(\kk'+2r\pp_F) \right| \le\cr
&C\int_0^{t_0}{dk'\over2\p} \int_0^\io {dk_0\over \sqrt{k_0^2+(v_0 k')^2}
\sqrt{k_0^2+v_0^4}} \le {C\over v_0}\; .\cr}\Eq(4.4) $$
The bound \equ(4.4) can be improved for $\s\to 0$, if $\o'=-\o$; we have
$$\eqalign{
&{1\over L}\sum_{k'\in \DD'_L}\int_{-\io}^{\io} {dk_0\over2\p}
\left| \tilde g_{\o,-\o}^{(0)}(\kk') \,
\tilde g^{(1)}_{1,1}(\kk'+2r\pp_F) \right| \le\cr
&C|\s|\int_0^{t_0}{dk'\over2\p}\int_0^\io {dk_0\over [k_0^2+(v_0 k')^2+\s^2]
\sqrt{k_0^2+v_0^4}} \le {C|\s|\over v_0^3}
\left(1+\log {v_0^2\over |\s|}\right) \; .\cr}\Eq(4.5)$$
Let us now consider the case $h=h'=0$; for any $\o_i,\o_i'$, we get
$$\eqalign{
&{1\over L}\sum_{k'\in \DD'_L}\int_{-\io}^{\io} {dk_0\over2\p}
\left| \tilde g_{\o_1,\o_1'}^{(0)}(\kk')\,
\tilde g_{\o_2,\o_2'}^{(0)}(\kk'+2r\pp_F) \right|\le\cr
&C\int_{-\p}^{\p} dk' \int_{-\io}^{\io} dk_0
{f_0(k')\over \sqrt{k_0^2+(v_0 k')^2+\s^2}}
{f_0(k'+2rp_F)\over \sqrt{k_0^2+v_0^2(k'+2rp_F)^2+\s^2}} \le\cr
&C\int_0^{t_0} dk' \int_0^{\io} {dk_0\over k_0^2+(v_0 k')^2+\s^2}
\le {C\over v_0} \left(
1+\log {v_0^2\over |\s|} \right) \; .\cr}\Eq(4.6)$$
If $h=h'=0$ and $\o_1\not=\o'_1$ or $\o_2\not=\o'_2$,
the last bound can be improved; in fact we get
$$\eqalign{
&{1\over L}\sum_{k'\in \DD'_L}\int_{-\io}^{\io} {dk_0\over2\p}
\left| \tilde g_{\o_1,\o_1'}^{(0)}(\kk')\,
\tilde g_{\o_2,\o_2'}^{(0)}(\kk'+2r\pp_F) \right|\le \cr
&C|\s|\int_0^{t_0} dk' \int_0^{\io} {dk_0\over [k_0^2+(v_0 k')^2+\s^2]^{3/2}}
\le {C\over v_0}\; .\quad \cr} \Eq(4.7) $$

\*

The previous bound can be further improved, if we suppose also that $r\not=0$,
by taking into account that, in this case, $\max\{|k'|,|k'+2rp_F|\}\ge \p/Q$.
Let us suppose, for example, that $\o_2=-\o'_2$; we have
$$\eqalign{
&{1\over L}\sum_{k'\in \DD'_L}\int_{-\io}^{\io} {dk_0\over2\p}
\left| \tilde g_{\o_1,\o_1'}^{(0)}(\kk')\,
\tilde g_{\o_2,-\o_2}^{(0)}(\kk'+2r\pp_F) \right|\le \cr
&C|\s|\int_{-\p}^{\p} dk' \int_0^{\io} dk_0
{f_0(k')\over \sqrt{k_0^2+(v_0 k')^2+\s^2}}
{f_0(k'+2rp_F)\over [k_0^2+v_0^2(k'+2rp_F)^2+\s^2]}
\le \cr
& {C|\s|Q\over v_0} \int_0^{t_0} dk' \int_0^{\io} {dk_0\over k_0^2+(v_0 k')^2+
\s^2}\le {C|\s|Q\over v_0^2} \left(1+\log {v_0^2\over |\s|}\right)
\; . \cr} \Eq(4.51)$$

\*

In the following we shall need also the bounds of the expression obtained
substituting in the r.h.s. of \equ(4.1) one of the two propagators with its
derivative with respect to $\s$. By proceeding as before, one can easily prove
that, for any $\o$ and any integer $r$,
$${1\over L}\sum_{k'\in \DD'_L}\int_{-\io}^{\io} {dk_0\over2\p}
\left| {\dpr \tilde g_{\o,\o}^{(0)}(\kk')\over \dpr\s} \,
\tilde g^{(1)}_{1,1}(\kk'+2r\pp_F) \right| \le
{C\over v_0^3}\; ,\Eq(4.7a)$$
$${1\over L}\sum_{k'\in \DD'_L}\int_{-\io}^{\io} {dk_0\over2\p}
\left| {\dpr \tilde g_{\o,-\o}^{(0)}(\kk')\over \dpr\s} \,
\tilde g^{(1)}_{1,1}(\kk'+2r\pp_F) \right| \le
{C\over v_0^3}\left(1+\log {v_0^2\over |\s|}\right)\; ;\Eq(4.7b)$$
that, for any $\o_i, \o'_i$ and any integer $r$,
$${1\over L}\sum_{k'\in \DD'_L}\int_{-\io}^{\io} {dk_0\over2\p}
\left| {\dpr \tilde g_{\o_1,\o'_1}^{(0)}(\kk')\over \dpr\s} \,
\tilde g_{\o_2,\o'_2}^{(0)}(\kk'+2r\pp_F) \right| \le
{C\over v_0|\s|}\; ,\Eq(4.7c)$$
and finally that, for any $\o_1,\o'_1,\o$ and any integer $r\not=0$,
$${1\over L}\sum_{k'\in \DD'_L}\int_{-\io}^{\io} {dk_0\over2\p}
\left| {\dpr \tilde g_{\o_1,\o'_1}^{(0)}(\kk')\over \dpr\s} \,
\tilde g_{\o,-\o}^{(0)}(\kk'+2r\pp_F) \right| \le
{C Q\over v_0^2}\; ,\Eq(4.7d)$$
$${1\over L}\sum_{k'\in \DD'_L}\int_{-\io}^{\io} {dk_0\over2\p}
\left| \tilde g_{\o_1,\o'_1}^{(0)}(\kk')\,
{\dpr \tilde g_{\o,-\o}^{(0)}\over \dpr\s}(\kk'+2r\pp_F) \right| \le
{C Q\over v_0^2}\; .\Eq(4.7h)$$

\*

\sub(4.1a) {\cs Remark.}
All the previous bounds are valid also if we exchange in the l.h.s. $\kk'$
with $\kk'+2r\pp_F$; this immediately follows from the observation that
the variable $k'$ is defined modulo $2\pi$.

\*

\sub(4.2)
We shall now consider the graphs contributing to the constants $c_n(\s)$
introduced in \equ(2.1), in order to prove Lemma \secc(2.6). We can write
$$-\l\hf{n}\,c_n(\s) = \sum_{\th\in\TT_{n,1}} \d_{n,n_v} \Val(\th) \; .
\Eq(4.8) $$
The equations \equ(4.8), \equ(4.1) and \equ(4.2) imply, if $|n|>2$,
$$\eqalignno{
c_n(\s) = & {1\over L}\sum_{k'\in \DD'_L} \int_{-\io}^{\io} {dk_0\over 2\p}
\Big\{
\tilde g_{1,1}^{(1)}(\kk')\,\tilde g_{1,1}^{(1)}(\kk'+2n\pp_F)\cr
+ \sum_{\o=\pm 1} & \Big[
\tilde g_{1,1}^{(1)}(\kk')\,\tilde g_{\o,\o}^{(0)}(\kk'+2n\pp_F+(1-\o)\pp_F)
& \eq(4.9) \cr
+ & \tilde g_{\o,\o}^{(0)}(\kk')\,\tilde g_{1,1}^{(1)}
(\kk'+2n\pp_F-(1-\o)\pp_F)
+ \tilde g_{\o,-\o}^{(0)}(\kk')\,\tilde g_{-\o,\o}^{(0)}(\kk'+2n\pp_F) \cr
+ &
\tilde g_{\o,\o}^{(0)}(\kk')\,\tilde g_{\o,\o}^{(0)}(\kk'+2n\pp_F) +
\tilde g_{\o,\o}^{(0)}(\kk')\,\tilde g_{-\o,-\o}^{(0)}(\kk'+(2n+2\o)\pp_F)
\Big]\Big\} \; . \cr} $$

By using the bounds \equ(4.3), \equ(4.4) and \equ(4.7), we see that the
first four terms in the r.h.s. of \equ(4.9) are bounded by $(C/v_0)(1+
\log v_0^{-1})$. However, the remaining terms, \ie those with
$h=h'=0$ and $\o_1-\o_1'=\o_2-\o_2'=0$, need a more careful analysis;
these terms will be denoted as
$$ a_n \= {1\over L}\sum_{k'\in \DD'_L}
\int_{-\io}^{\io} {dk_0\over2\p}
\, \tilde g_{\o,\o}^{(0)}(\kk')\,
\tilde g_{\o,\o}^{(0)}(\kk'+2n\pp_F) \; , \Eq(4.10) $$
when $\o_1=\o_2$, and
$$ b_{n,\o} = {1\over L}\sum_{k'\in \DD'_L}
\int_{-\io}^{\io} {dk_0\over2\p}
\, \tilde g_{\o,\o}^{(0)}(\kk')\,
\tilde g_{-\o,-\o}^{(0)}(\kk'+(2n+2\o)\pp_F) \; . \Eq(4.11) $$
when $\o_1=-\o_2$. The following two Lemmata \secc(4.4) and \secc(4.6)
show that the dimensional bounds which would follow from \equ(4.6)
in fact can be improved.

\*

\sub(4.3) {\cs Remark.} Note that $a_n$ is a $\o$-independent
quantity, so that we can set $\o=1$ in \equ(4.10); this property easily
follows from the observation that $g^{(0)}_{\o,\o}(k',k_0)=
g^{(0)}_{-\o,-\o}(-k',k_0)$, see \equ(3.23). It is also easy
to prove that $b_{n,1}=b_{-n,-1}$.

\*

\sub(4.4) {\cs Lemma.}
{\it Let $|n|\ge 2$ and let $a_n$ be defined as in \equ(4.10); then
$|a_n|<C/v_0$.}

\*

\sub(4.5) {\it Proof of Lemma \secc(4.4).}
By Remark \secc(4.3), it is enough to study the case $\o=1$ in \equ(4.10).
Define
$$\eqalign{
\bar g^{(0)}_{\o,\o}(\kk') &=
{f_0(k')\over -ik_0 +\FF(\o k') } \; ,\cr
\FF(k') &\= \sign(k')\sqrt{ F_2^2(k')+ \s^2_0(k')} -F_1(k') \; ,}\Eq(4.12)$$
and
$$\eqalign{
\bar a_n &\= {1\over L}\sum_{k'\in \DD'_L}
\int_{-\io}^{\io} {dk_0\over2\p}
\bar g^{(0)}_{1,1}(\kk') \bar g^{(0)}_{1,1}(\kk'+2n\pp_F)\cr
&= {1\over L}\sum_{k'\in \DD'_L}
\; f_0(k')\,f_0(k'+2np_F) \, \AA_n(k')\;.\cr}\Eq(4.13) $$
where $\AA_n(k')$ is obtained by explicitly performing
the integral on $k_0$. It is easy to see that, defining
$$ s(k') = \sign\Big(\FF(k')\Big)\;, \Eq(4.14) $$
if $s(k')=s(k'+2np_F)$, one has $ \AA_n(k') =0$,
while, if $s(k')=-s(k'+2np_F)$, one has
$$\AA_n(k') = s(k') \Big[\FF(k'+2np_F)-\FF(k')\Big]^{-1} \; .\Eq(4.15)$$
Note that, by \equ(3.26), $s(k')=\sign(k')$, if $|k'|\le t_0$, \ie on the
support of $f_0(k')$; hence we have
$$\bar a_n = -{1\over L}\sum_{k'\in \DD'_L\cap\DD'_*}
\; { f_0(k')\,f_0(k'+2np_F)
\over |\FF(k')| + |\FF(k'+2np_F)| } \; ,\Eq(4.16)$$
where
$$ \DD'_* = \{ k'\in[-t_0,t_0]\; : \;\sign(k')=-\sign(k'+2np_F)\}\; .
\Eq(4.17)$$

We want to show that
$$\max\{ |\FF(k')|,|\FF(k'+2np_F)|\}
\ge {c_2\over 2} \|2np_F\|_{\ttt_1} \= \D_1 \; , \Eq(4.18)$$
if $k'\in \DD'_*$, $k'+2np_F \in [-t_0,t_0]$ and $c_2=(\sqrt{2}/\p)v_0$.

If $|\FF(k')|\ge \D_1$, \equ(4.18) is immediately verified. Let us suppose
now that $|\FF(k')|<\D_1$; then, by using \equ(3.26), we get
$$ |\FF(k')| \ge |F_2(k')| - |F_1(k')| \ge {1\over 2}|F_2(k')|
> c_2|k'|\; , \Eq(4.19)$$
so that
$$ |k'| < {\D_1\over c_2} = {1\over2} \|2np_F\|_{\ttt_1}\; ,\Eq(4.20)$$
implying
$$\|k'+2np_F\|_{\ttt_1} \ge \Big| \|2np_F\|_{\ttt_1}-|k'| \Big|
\ge {1\over2} \|2np_F\|_{\ttt_1}\; .\Eq(4.21)$$
Moreover, since $\|k'+2np_F\|_{\ttt_1}\le t_0$, then
$$\left|\FF(k'+2np_F)\right| > c_2 \|k'+2np_F\|_{\ttt_1}\; ;\Eq(4.22)$$
hence, by using \equ(4.21) and \equ(4.22), we get
$$\left|\FF(k'+2np_F)\right| \ge {c_2\over 2} \|2np_F\|_{\ttt_1} = \D_1 \;,
\Eq(4.23)$$
which implies \equ(4.18) also when $|\FF(k')|<\D_1$.

Inserting \equ(4.18) into \equ(4.16) leads to
$$|\bar a_n| \le \int_{\DD'_*} {dk' \over \p c_2 \|2np_F\|_{\ttt_1}}
\le {\sqrt{2}\over v_0}\; , \Eq(4.24)$$
as the size of the set $\DD'_*$ is bounded by $2\|2np_F\|_{\ttt_1}$.

\*

In order to complete the proof of Lemma \secc(4.4), we note that
$$ \eqalign{
a_n - \bar a_n & = {1\over L}\sum_{k'\in \DD'_L}
\int_{-\io}^{\io} {dk_0\over2\p}
\Big\{ \big[ \tilde g^{(0)}_{1,1}(\kk') -
\bar g^{(0)}_{1,1}(\kk')  \big] \tilde g^{(0)}_{1,1}(\kk'+2n\pp_F) \cr
& + \bar g^{(0)}_{1,1}(\kk') \big[
\tilde g^{(0)}_{1,1}(\kk'+2n\pp_F) - \bar g^{(0)}_{1,1}(\kk'+2n\pp_F)
\big] \Big\} \; .\cr} \Eq(4.25)$$
Moreover, by \equ(3.23) and \equ(4.12),
$$|\tilde g^{(0)}_{1,1}(\kk') -\bar g^{(0)}_{1,1}(\kk')| =
{\sqrt{F_2(k')^2+ \s^2_0(k')} - |F_2(k')| \over
\big| [-ik_0 - F_1(k')]^2 - [ F_2^2(k') + \s^2_0(k')] \big|} |f_0(k')|
\; ,\Eq(4.26)$$
so that, by using also \equ(3.26), we get
$$|a_n - \bar a_n| \le C \int_0^{t_0} dk'\int_0^{\io} dk_0
{\s^2 \over \sqrt{ (v_0k')^2 + \s^2} \; [k_0^2+(v_0k')^2+\s^2]^{3/2}}
\le {C\over v_0}\;.\Eq(4.27)$$
The bounds \equ(4.24) and \equ(4.27) imply Lemma \secc(4.4). \qed

\*

\sub(4.6) {\cs Lemma.}
{\it Let $|n|\ge 2$, and let $ b_{n,\o}$ be defined as in \equ(4.11);
then $|b_{n,\o}| \le (C\log Q)/v_0$.}

\*

\sub(4.7) {\it Proof of Lemma \secc(4.6).}
Let us define $\bar b_{n,\o}$ as
$$\bar b_{n,\o} =
{1\over L}\sum_{k'\in \DD'_L}\int_{-\io}^{\io} {dk_0\over2\p}
\; \bar g^{(0)}_{\o,\o}(\kk')\,\bar g^{(0)}_{-\o,-\o}(\k_\o',k_0)
\; , \Eq(4.28) $$
where $\k_\o'=k'+(2n+2\o)p_F$ and $\bar g^{(0)}_{\o,\o}(\kk')$
is defined in \equ(4.12), and set $b_{n,\o}=\bar b_{n,\o}+\II'$.
By dimensional bounds analogous to those which led to \equ(4.27), it
is easy to prove that $|\II'|\le C/v_0$.
Moreover, by proceeding as in \sec(4.5), we see that
$$\bar b_{n,\o} = -{1\over L}\sum_{k'\in \DD'_L\cap\DD'_\o}\;
{ f_0(k')\,f_0(\k_\o')
\over |\FF(\o k')| + |\FF(-\o \k_\o')| } \; ,\Eq(4.29)$$
where $\DD'_\o = \{ k'\in[-t_0,t_0] \; : \;\sign(\o k')=\sign(\o \k_\o')\}$.

By using the bound \equ(4.19), we have
$$|\FF(\o k')| + |\FF(-\o \k_\o')| \ge c_2 (|k'|+\|\k'_\o\|_{\ttt_1})
\Eq(4.30) $$
Moreover, since $|n|\ge 2$, $\|2np_F\pm 2p_F\|_{\ttt_1} \ge {2\p\over Q}$;
hence
$$|\bar b_{n,\o}| \le \int_{\DD'_\o} {dk'\over 2\p c_2
(|k'|+\|\k_\o'\|_{\ttt_1})} \le {C \log Q\over v_0} \; .\Eq(4.31)
$$
This completes the proof of Lemma \secc(4.6). \qed

\*

\sub(4.8) {\it Proof of Lemma \secc(2.6).} The bound \equ(2.6)
immediately follows from
the remark after \equ(4.9), Lemma \secc(4.4) and Lemma \secc(4.6).
The bound \equ(2.7) is easily proven from \equ(4.9)
by using the bounds \equ(4.7a)$\div$\equ(4.7c). \qed

\*
\sub(4.9) {\it Proof of Lemma \secc(2.13a).}
The definition of $c_1(\s)$ in \sec(2.3), \equ(3.32), \equ(4.1) and \equ(4.2)
imply that
$$\eqalignno{
c_1(\s) &= {1\over L}\sum_{k'\in \DD'_L} \int_{-\io}^{\io} {dk_0\over 2\p}
\Big\{ {\tilde g_{-1,1}^{(0)}(\kk') \over \s} +
\tilde g_{1,1}^{(1)}(\kk')\,\tilde g_{1,1}^{(1)}(\kk'+2\pp_F)\cr
&+ \sum_{\o=\pm 1} \Big[
\tilde g_{1,1}^{(1)}(\kk')\,\tilde g_{\o,\o}^{(0)}(\kk'+(3-\o)\pp_F) +
\tilde g_{\o,\o}^{(0)}(\kk')\,\tilde g_{1,1}^{(1)}(\kk'+(1+\o)\pp_F)
\Big]\Big\} \cr
&\= -F(\s,L)+\tilde c_1(\s)\;, & \eq(4.32) \cr} $$
where $-F(\s,L)$ denotes the first term in the r.h.s. of \equ(4.32), while
$\tilde c_1(\s)$ is the sum of the other ones. It turns out that $-F(\s,L)$ is
the leading term for $\s\to 0$; moreover it is the only term whose dependence
on $L$ is not trivial, hence we decide to indicate it explicitly.

By using the definition \equ(3.23) and by performing the integration over
$k_0$, we get
$$F(\s,L)= {1\over 2L}\sum_{k'\in \DD'_L} {f_0(k')^2\over
\sqrt{v_0^2 \sin^2 k'+\s^2 f_0(k')^2}}\;.\Eq(4.33)$$
The definition \equ(4.1a) of $\DD'_L$ implies that, for any finite volume
$L\=L_i$, the r.h.s. is singular for $\s\to 0$,
only if $\d=0$, that is only if
the number of Fermions is even; in that case, in fact, $k'=0$ belongs to the
set $\DD'_L$. It follows that, if $\d=1$, the equation
\equ(2.5) has no solution for $\l^2$ very small, how small depending on $L$;
this is the main source of the lower bound on $\l$ of Theorem \secc(1.7).
[Equivalently, for fixed $\l$ verifying the inequality to the right
in \equ(1.16), which is uniform in $L$, $L$ has to be large enough
so that also the inequality to the left in \equ(1.16) can be fulfilled.]
We separate the term with $k'=0$, if it is present, by writing
$$F(\s,L)= {1-\d\over 2L\s} + F_0(\s,L)\;.\Eq(4.34)$$
It is easy to see that
$$F_0(\s,L) = F_1(\s,L) + d_1(\s,L) + d_2(\s,L)\;,\Eq(4.35)$$
where
$$F_1(\s,L)= \sum_{n\not= 0:|2\p L^{-1}(n+\d/2)| \le t_0/2} {1\over
\sqrt{(2\p v_0)^2 (n+\d/2)^2 + (\s L)^2 }}\;,\Eq(4.36)$$
$$d_1(\s,L)= {1\over 2L}\sum_{k'\in\DD'_L \atop
|k'| \ge t_0/2} {f_0(k')^2\over
\sqrt{v_0^2 \sin^2 k'+\s^2 f_0(k')^2}}\;,\Eq(4.37)$$
$$d_2(\s,L)= {1\over 2L}\sum_{k'\in\DD'_L \atop
0\not= |k'| \le t_0/2} \left[
{1\over \sqrt{v_0^2 \sin^2 k'+\s^2}}
-{1\over \sqrt{(v_0 k')^2+\s^2}} \right]\;.\Eq(4.38)$$

Note that the sum in the r.h.s. of \equ(4.36) is empty, if
$[t_0 L/(4\p) + \d/2]<1$; in that case the equation \equ(2.5) may have a
solution, for $\l$ small enough, only if $\d=0$.
Hence we shall suppose that:
$$t_0 L \ge 4\p\;,\Eq(4.39)$$
a condition which is certainly verified, if the conditions \equ(2.18a) are
satisfied, since
$${4\over \p} \le v_0/t_0 \le 2\;.\Eq(4.40)$$.

By using \equ(4.40) and supposing that
$${|\s|\over v_0^2} \le 1\;,\Eq(4.41)$$
it is easy to show that
$$\sum_{i=1}^2 |d_i(\s)| \le {C\over v_0}\;,\qquad
\sum_{i=1}^2 \left|{\dpr d_i(\s)\over \dpr\s}\right| \le {C\over v_0^3}\;.
\Eq(4.42)$$

By substituting the sum in the r.h.s. of \equ(4.36) with an integral, we can
write
$$F_1(\s,L)=F_2(\s,L)+d_3(\s,L)\;,\Eq(4.43)$$
where
$$F_2(\s,L)= \int_{1-\d/2}^{t_0 L/(4\p)}\;{dx\over
\sqrt{(2\p v_0 x)^2 + (\s L)^2 }}\;.\Eq(4.44)$$
It is easy to see that, if the condition \equ(4.41) is verified, together with
$${v_0\over L|\s|} \le \tilde\e \le 1\;,\Eq(4.45)$$
then
$$|d_3(\s)| \le {C\over v_0}\;,\qquad
\left|{\dpr d_3(\s)\over \dpr\s}\right| \le {C\tilde\e\over v_0|\s|}\;.
\Eq(4.46)$$

The integral defining $F_2(\s,L)$ can be explicitly calculated; we get
$$F_2(\s,L)= {1\over 2\p v_0} \log {
{v_0 t_0\over 2|\s|} + \sqrt{\left({v_0 t_0\over 2\s}\right)^2 + 1}
\over {2\p v_0\over L|\s|}\left(1-{\d\over 2}\right) +
\sqrt{\left({2\p v_0\over L|\s|}\right)^2
\left(1-{\d\over 2}\right)^2 +1}  }\;.\Eq(4.47)$$
If we write
$$F_2(\s,L)= {1\over 2\p v_0} \log {v_0^2\over |\s|} + d_4(\s,L)\;,\Eq(4.48)$$
it is easy to prove, using \equ(4.40), \equ(4.41) and \equ(4.45), that
$$|d_4(\s)| \le {C\over v_0}\;,\qquad
\left|{\dpr d_4(\s)\over \dpr\s}\right| \le {C\over v_0}
\left(1+{\tilde\e\over |\s|}\right)\;.\Eq(4.49)$$

Finally, the function $\tilde c_1(\s)$ introduced in \equ(4.32) and its
derivative can be bounded, by using \equ(4.3), \equ(4.4) and \equ(4.7a), as
$$|\tilde c_1(\s)| \le {C\over v_0}\left(1+\log {1\over v_0} \right) \;,\qquad
\left|{\dpr\tilde c_1(\s)\over \dpr\s}\right| \le {C\over v_0^3}\;.
\Eq(4.49a)$$

It is now sufficient to define
$$r_1(\s)=2\p v_0 \left[ \sum_{i=1}^4 d_i(\s,L) + {1-\d\over 2L\s}
-\tilde c_1(\s) \right]\;,\Eq(4.50)$$
to complete the proof of Lemma \secc(2.13a). \qed

\pagina
\section(5,Bounds on the density perturbative expansion)

\sub(5.1) In this section we give some bounds
about the perturbative expansion \equ(3.34) of the function
$\tilde\r_n(\s,\F)$, in\-tro\-duced in \equ(2.1),
and we prove Lemmata \secc(2.2), \secc(2.9), \secc(2.10) and \secc(2.15).

Given $\F\in\FF$, let us define
$$R(\F)_n^{(q)}(\s) = \tilde\r_n^q(\s,\F(\s))\;,\quad |n|>0, \quad
q>0\;.\Eq(5.1)$$
Moreover, if ${\cal J}$ is the space of the
$C^1$-functions of $\s\in J$ with values in $\RRR$,
and $r(\s)\in {\cal J}$,
we shall define, in agreement with \equ(2.8),
$$\|r\|_{{\cal J}} \= \sup_{\s\in J} \left[ |\s|^{-1} |r(\s)|
+\left|{\dpr r\over \dpr\s}(\s)\right|\right] \;.\Eq(5.1a)$$

\*

\sub(5.2) {\cs Lemma.} {\it If $\F\in\BB$ and $|\s|\le v_0^2$,
then, for any $n\neq 0$ and $q>0$,
$$ \eqalign{
& \qquad \|R(\F)_n^{(q)}\|_{{\cal J}} \le \cr
& Dv_0\left(1+\log{1\over v_0}\right) \left( {C \over v_0^2}
\right)^q q^2 \left({3q\over |n|}\right)^N \left[1+(1-\d_{q,1})
\log {v_0^2\over |\s|}\right] (|\s|Q)^{[q/2]} \; , \cr} \Eq(5.2)$$
where $C$ and $D$ are suitable constants.}

\*

\sub(5.3) {\it Proof of Lemma {\secc(5.2)}.} In order to bound
$\tilde\r_n^{q}(\s,\F)$, we shall use the expansion in \equ(3.34).
Let $\th\in\TT_{n,q}$ be one of the graphs contributing
to $\tilde\r_n^{q}(\s,\F)$ and $v$ one of its vertices.
If $|n_v|\neq 1$, one has (see \sec(3) for notations)
$$ \|k'_{\ell_v}-k'_{\ell_{v'}}\|_{\ttt^1} =\| 2n_v p+(\o^2_{\ell_v}-
\o^1_{\ell_{v'}})p\|_{\ttt^1} \ge {2\p\over Q} \; ,\Eq(5.4) $$
so that
$$ \max \{ \|k'_{\ell_v}\|_{\ttt^1},\|k'_{\ell_{v'}}\|_{\ttt^1} \}
\ge {\p\over Q} \; .\Eq(5.5) $$
Then there is a constant $C_2$ such that, $\forall v \in \th$, $|n_v|\not=1$,
if $|\s|\le 1$,
$$\eqalignno{
\left| \tg{\ell_v} \tg{\ell_{v'}} \right| &\le {C_2 Q\over v_0^4|\s|}\; ,&
\eq(5.6)\cr
\min \left\{ |\tg{\ell_v}| ,|\tg{\ell_{v'}}| \right\} &\le {C_2 Q\over v_0^2}
\;,&\eq(5.7)\cr}$$
by \equ(5.5), \equ(3.28) and \equ(3.27), since $v_0\le 1$.

Note that \equ(5.6) and \equ(5.7) still hold for $|n_v|=1$, as,
in such a case,
$h_{\ell_v}=h_{\ell_{v'}}=0$ is not allowed (see \sec(3.7)) and
the support properties imply that both propagators are bounded by $C/v_0^2$.

Note also that, thanks to \equ(3.31),
$$|n| \le q+1 +\sum_v |n_v| \le 3 \sum_v |n_v| \qquad \Rightarrow
\qquad \exists v^*\; : \; |n_{v^*}| \ge {|n|\over 3q}\; .\Eq(5.8)$$

Let us now suppose that $q=2\bq$, with $\bq\ge1$. It is easy to see that, in
this case, it is possible to couple $2\bq$ among the $2\bq+1$ propagators
appearing in the expression of $\Val(\th)$, see \equ(3.33), in $\bq$ pairs
$\left\{ \tg{\ell_v}, \tg{\ell_{v'}} \right\}$ with $v\not=v^*$; let
$\tg{\ell^{(1)}}$, $\ell^{(1)}=\ell_{v^*}$,
the propagator left alone after this coupling operation.
We select in an arbitrary way one of the $\bq$ couples and we use the
bound \equ(5.7) for one of the propagators belonging to it; let
$\tg{\ell^{(2)}}$ the other propagator of the selected couple. The propagators
of all the other couples will be bounded by \equ(5.6). We get
$$ |\Val(\th)| \le (C_2 Q)^{\bq} {|\s|^{-\bq+1}\over v_0^{4\bq-2}}
\Big( \prod_{i=1}^{2\bq} |\F_{n_{v_i}}| \Big)
{1\over L\b} \sum_{\kk'\in {\cal D}_{L,\b}}
|\tg{\ell^{(1)}} \tg{\ell^{(2)}}|\;.\Eq(5.9) $$

Let us now suppose that $|\s|\le v_0^2$; then we can use the bounds
\equ(4.3)--\equ(4.7), valid also for finite $\b$, to prove that,
for any choice of $\tg{\ell^{(1)}}$ and $\tg{\ell^{(2)}}$,
$${1\over L\b} \sum_{\kk'\in {\cal D}_{L,\b}}
\, |\tg{\ell^{(1)}} \tg{\ell^{(2)}}|\le
{D_2\over v_0}\left(1+\log{1\over v_0}\right)
\left(1+\log{v_0^2\over |\s|}\right)\;.\Eq(5.10) $$
Hence, if $\F\in\BB$, by using \equ(2.9), \equ(5.8) and \equ(5.10), we get
$$ \eqalign{
\sum_{\th\in\TT_{n,2\bq}} \left| \Val(\th) \right| & \le
{D_2\over v_0^{2q-1}} \left(1+\log{1\over v_0}\right) \cr
& D_1^q q
C_2^{\bq} |\s| (|\s| Q)^{\bar q} \left(1+\log
{v_0^2\over |\s|}\right)\left( {3q\over |n|} \right)^N \;,\cr} \Eq(5.11)$$
where $D_1^qq$ takes into account the fact that
there are $5$ possible choices for the
$\o_\ell^{1},\o_\ell^{2},h_\ell$ labels for each line, and
$q$ possible choice for the vertex $v^*$;
then the bound \equ(5.2) is proved for even $q$.

The case $q=2\bq+1$, with $\bq\ge1$, can be treated in a similar way. We note
that it is always possible to couple $2\bq$ among the $2\bq+2$ propagators
appearing in the expression of $\Val(\th)$ in $\bq$ pairs
$\left\{ \tg{\ell_v} , \tg{\ell_{v'}} \right\}$ with $v\not=v^*$; let
$\tg{\ell^{(1)}}$ and $\tg{\ell^{(2)}}$ be
the propagators left alone after this coupling operation.
Then we use \equ(5.6) for all the couples and the bound \equ(5.10) for the two
remaining propagators. We get a bound similar to \equ(5.9), with $|\s|^{-\bq}$
in place of $|\s|^{-\bq+1}$, but the final bound is the same as before.

We still have to consider the case $q=1$. We could of course get again the
previous bound with $\bq=0$, but there is now an improvement,
which will play an important role.
The improvement follows from the observation that, if
$q=1$, the graphs contributing to $\tilde\r^1_n$ have only
one vertex with Fourier index $n_{v_1}\not=n$,
so that at least one of the two propagators must
have different $\o$-indices. By using the bound \equ(4.7),
this implies that the bound \equ(5.10) can be improved by erasing
the factor $[1+\log (v_0^2/|\s|)]$.

In order to complete the proof of \equ(5.2), we have to bound also
\hfill\break
$\dpr\tilde\r_n^{q}(\s,\F(\s))/$ $\dpr\s$.
We can proceed as before, by noticing
that $\dpr \Val(\theta)/\dpr\s$ can be
written as the sum of $2q+1$ terms, each term differing
from $\Val(\theta)$ only because there is the derivative acting
on a single propagator or a single vertex function.
If the derivative acts on one of the coupled propagators, one can use the
bounds \equ(5.6) and \equ(5.7) modified so that the r.h.s.
is multiplied by $|\s|^{-1}$; if the derivative acts on a
vertex function, since $\F\in\BB$, one can use the bound
$|\dpr\F_n(\s)/\dpr\s|\le |n|^{-N}$; if the derivative acts
on one of the propagators left alone after the coupling operation, one can
use the bound, following from \equ(4.7a)$\div$\equ(4.7c), if $|\s|\le v_0^2$,
$$ {1\over L\b} \sum_{\kk'\in {\cal D}_{L,\b}}
\, \left| {\dpr\tg{\ell^{(1)}}\over \dpr\s} \tg{\ell^{(2)}}\right|\le
{D_3\over v_0|\s|}\;.\Eq(5.10a)$$
We get, for any $q>0$,
$$ \eqalign{
\left| {\dpr R(\F)_n^{(q)} \over \dpr \s} (\s) \right| \le &
\sum_{\th\in\TT_{n,2\bq}} \left| {\dpr\Val(\th) \over\dpr\s}\right| \le
\cr & (2q+1) {D_3\over v_0^{2q-1}}  D_1^q q
C_2^{\bq} (|\s| Q)^{\bar q} \left( {3q\over |n|} \right)^N \;,
\cr} \Eq(5.11a)$$
with $\bar q=[q/2]$. This complete the proof of \equ(5.2). \qed

\*

\sub(5.4) {\it Proof of Lemma \secc(2.2).}
The bound \equ(5.2) immediately implies that $\tilde\r_n^q(\s,\F)$ is summable
over $q$, for $\s Q/v_0^4$ small enough, uniformly in $i$, $\r$ and $\b$.
On the other end,
it is easy to see that the bound is valid also if we substitute in the
expression \equ(3.33) of $\Val(\th)$ the sum over $k_0$ with the integral on
the real axis and that $\lim_{\b\to\io} \sum_{q=1}^\io \tilde\r_n^q(\s,\F)$
is obtained from $\sum_{q=1}^\io \tilde\r_n^q(\s,\F)$ by doing this
substitution.
The claim of the Lemma about the continuous dependence on $\l$ of
$\hat\r_n(\f,\mu)$ is an easy consequence of this remark and \equ(3.32).

In a similar way, one can see that $\lim_{i\to\io}\hat\r_n(\f,\mu)$ is
obtained by substituting in the expression of $\Val(\th)$ the sum over $k'$
with an integral over the interval $[-\p,\p]$ and that this limit is also
continuous in $\l$ near $0$.

The other claims of the Lemma about the density and the gap of $\bf h$ can be
proved as in [BGM], \S4.5 and \S4.6. In [BGM] a more complicated expansion was
used (involving a further decomposition of the field $\psi^{(0)}$), but the
proof of these two points can be even more simply reached by using the
expansion of this paper and bounds of the graphs similar to \equ(5.11).
We shall not give here the details, but we only remark that the main point in
the proof is the remark that the propagators \equ(3.23) are analytic in $k_0$
in the strip $|\Im k_0|\le |\s|/2$. \qed

\*

\sub(5.5) {\it Proof of Lemma \secc(2.9).} By the remark in \sec(5.4),
$\lim_{\b\to\io}$ can be exchanged with the sum over $q$ in \equ(3.32) and
\equ(3.34). In the following, for simplicity, we shall use the notation
$\tilde\r_n^q(\s,\F)$ to identify $\lim_{\b\to\io}\tilde\r_n^q(\s,\F)$.

Let us suppose that $\F,\F'\in \BB$ and $q\ge 2$; then, by \equ(3.32)
$$\tilde\r_n^{q}(\s,\F')-\tilde\r_n^{q}(\s,\F) = \sum_{\th\in\TT_{n,q}
(\F,\F')}\Val(\th) \; ,\Eq(5.12)$$
where $\TT_{n,q}(\F,\F')$ is a set of labeled graphs whose definition differs
from the definition of $\TT_{n,q}$, see \sec(3.7), only because there is a
new label $\a_v\in\{0,1,2\}$ for each vertex; moreover
$$\Val(\th) = - {1\over L}\sum_{k'\in\DD'_L}\int_{-\io}^{\io}{dk_0\over 2\p}
\Big( \prod_{i=1}^{q+1} \tg{\ell_i} \Big)
\Big( \prod_{v\in\th} F_{n_v}^{\a_v} \Big) \; ,\Eq(5.13)$$
$$F_{n_v}^{\a_v} = \cases{ \s & if $\a_v=0\; ,$\cr \F_{n_v} & if $\a_v=1\; ,$
\cr \F'_{n_v}-\F_{n_v} & if $\a_v=2\; ,$\cr}\Eq(5.14)$$
and there is the constraint that at least one vertex has label $\a_v=2$.

By proceeding as in the proof of Lemma \secc(5.2) and using
the definitions \equ(2.8) and \equ(5.1), we get, if $q\ge 2$,
$$\eqalign{
&\qquad \|R(\F')^{(q)} - R(\F)^{(q)}\|_\FF \le \cr
&D\,v_0\left(1+\log{1\over v_0}\right) \left({C\over v_0^{2}}\right)^q
\|\F'-\F\|_\FF \, q^2 (3q)^N
\left(1+ \log {v_0^2\over |\s|}\right) (|\s| Q)^{[q/2]}\;.\cr}\Eq(5.15)$$
Hence, if $Q v_0^{-4}|\s|[1+ \log (v_0^2/|\s|)]\le 1$
and $|\s| Q v_0^{-4}\le 1/(2C^2)$, we have
$$\sum_{q=2}^\io \|R(\F')^{(q)} - R(\F)^{(q)}\|_\FF \le
{C_1\over v_0}\left(1+\log{1\over v_0}\right)
3^N N!\; \|\F'-\F\|_\FF\; .\Eq(5.16)$$
with a suitable constant $C_1$.

In order to complete the proof of the lemma, we have to estimate
$\|R(\F')^{(1)} - R(\F)^{(1)}\|_\FF$.
The bound \equ(5.15), with $q=1$, is
still valid, but it is not sufficient; however there is the improvement
with respect to \equ(5.15) due to the fact that, if $\th$ is a
graph contributing to $\tilde \r_n^{1}(\s,\F')-
\tilde \r_n^{1}(\s,\F)$, the only vertex
belonging to $\th$ has a Fourier index $n_{v_1}\not=n$.
As in the proof of Lemma \secc(5.2), this remark allows to eliminate the
factor $[1+\log (v_0^2/|\s|)]$ in the bound \equ(5.15) for any value of $n$.
The previous remark implies that
$$\|R(\F')^{(1)} - R(\F)^{(1)}\|_\FF \le
{C\over v_0}\left(1+\log{1\over v_0}\right)\|\F'-\F\|_\FF\; .\Eq(5.17)$$
This bound and \equ(5.16) immediately imply Lemma \secc(2.9). \qed

\*

\sub(5.6) {\it Proof of Lemma \secc(2.10).} The graph expansion of
$\tilde\r_n^q(\s,0)$ has the property that, given a graph
$\th\in\TT_{n,q}$ with $\Val(\th)\not=0$, each vertex of
$\th$ has Fourier index $n_v=\pm 1$.
This implies that, for any $v\in\th$ (see \sec(3.7)),
$h_{\ell_v}=h_{\ell_{v'}}=0$ is not allowed, so that the
number of non diagonal propagators is less or equal of $\bar q+1$,
if $\bq=[q/2]$. Hence \equ(3.31) implies that
$$ |n| \le q + \bar q +1 \; .\Eq(5.18) $$
We can bound $\Val(\th)\not=0$ as in \sec(5.3), by choosing in an arbitrary
way the vertex $v^*$ (since we do not need now to extract the factor
$|n|^{-N}$), and we get
$$ \sum_{\theta\in\TT_{n,q}} \left| \Val(\th) \right| \le {D\over v_0^{2q-1}}
\left(1+\log{1\over v_0}\right) 5^{q+1} C^{\bq} |\s| (|\s| Q)^{\bar q}
\left(1+ \log {v_0^2\over |\s|}\right) \; .\Eq(5.19) $$
It is easy to see that, if $q\ge 2$, $\bar q\ge |n|/5$; hence,
if $Qv_0^{-3}|\s|^{1/2} [1+ \log (v_0^2/|\s|)]$ is small enough,
$$\sum_{q=2}^\io ||R(0)_n^{(q)}||_{{\cal J}} \le {C\over v_0}
\left(1+\log{1\over v_0}\right)
\left({|\s|\over v_0^2}\right)^{|n|\over 10}\;.\Eq(5.20)$$

\*

In order to complete the proof of Lemma \secc(2.10), we have to
improve the bound \equ(5.19) in the case $q=1$. Note that
$\tilde \r_n^1(\s,0)$ is different from $0$ only if $|n|=2$ (only one
propagator may have frequency label $h=0$) and it
is given, if $n=2$ (the case $n=-2$ is similar), by
$$\s {1\over L}\sum_{k'\in\DD'_L} \int_{-\io}^{\io}{dk_0\over 2\p}
\big[ \tilde g^{(1)}_{1,1}(\kk') \, \tilde g^{(0)}_{-1,1}(\kk'+ 4\pp_F) +
\tilde g^{(0)}_{-1,1}(\kk') \, \tilde g^{(1)}_{1,1}(\kk'+ 2\pp_F) \big]
\;.\Eq(5.21) $$
Hence, by using \equ(4.5) and \equ(4.7b), we get
$$||R(0)_2^{(1)}||_{{\cal J}} \le
{C\over v_0^3}|\s|\left(1+ \log {v_0^2\over |\s|}\right) \le
{C\over v_0}\, \left({|\s|\over v_0^2}\right)^{2/10}\;.\Eq(5.22)$$
This bound and the bound \equ(5.20) imply \equ(2.11). \qed

\*
\sub(5.7) {\it Proof of Lemma \secc(2.13b).}
By Lemma \secc(5.2), if $Qv_0^{-3}|\s|^{1/2} [1+ \log (v_0^2/|\s|)]$
is small enough, which is certainly true if condition \equ(2.11a) is
satisfied, with $\e$ small enough, we have
$$\sum_{q\ge 2} |\tilde\r_1^q(\s,\F)| \le |\s|\;{C^N N!\over v_0}
\left(1+\log{1\over v_0}\right) \left({|\s|\over
v_0^2}\right)^{1/2}\;.\Eq(5.23)$$
Moreover, since the graphs contributing to $\tilde\r_1^1(\s,\F)$ have only one
vertex with index $n_v=\pm 2,\pm 3$, at least one of its two propagators has
frequency index $h=0$ and different $\o$-indices. It follows, by
\equ(4.5), \equ(4.7) and Lemma \secc(2.11), that
$$|\tilde\r_1^1(\s,\F)| \le {C\over v_0} (|\F_2(\s)|+|\F_3(\s)|) \le
{C\over v_0}|\s| \left({\l^2\over v_0}\right)^N\;.\Eq(5.24)$$
Hence, if $|\s|^{1/4}C^N N! (1+ \log v_0^{-1}) \le v_0^{1/2}$,
which is certainly
true if condition \equ(2.11a) is satisfied, with $\e$ small enough, and
$r_2(\s)$ is defined as in \equ(2.19b), we get
$$ \left| \tilde\r_1^1 (\s) \right| \le
C {|\s|\over v_0} \left[\left({|\s|\over v_0^2}\right)^{1/4}
+ \left({\l^2\over v_0}\right)^N \right] \; , \Eq(5.24a) $$
which implies the bound in the first line of \equ(2.19c).

Let us now consider the derivative of $r_2(\s)$.
By Lemma \secc(5.2), if $|\s|^{1/2} [1+ \log (v_0^2/|\s|)]\ Qv_0^{-3}$
is small enough, we have
$$\sum_{q\ge 2} \left|{\dpr\tilde\r_1^q(\s,\F(\s))\over \dpr\s} \right| \le
{C^N N!\over v_0} \left(1+\log{1\over v_0}\right) \left({|\s|\over
v_0^2}\right)^{1/2}\;.\Eq(5.25)$$
Moreover, by Lemma \secc(2.11) and the remark preceding \equ(5.24),
$$\left|{\dpr\tilde\r_1^1(\s,\F(\s))\over \dpr\s} \right| \le
{C\over v_0} \|\F\|_\FF \le
{C\over v_0}\left({\l^2\over v_0}\right)^N\;.\Eq(5.26)$$
It follows that, if $|\s|^{1/4}C^N N! (1+ \log v_0^{-1}) \le v_0^{1/2}$,
which is certainly true if condition \equ(2.11a)
is satisfied, with $\e$ small enough,
$$\left|{\dpr r_2\over \dpr\s}(\s) \right| \le
{2\p v_0\over |\s|} ||R(\F)_1||_{{\cal J}}  \le {C\over |\s|}
\left[\left({|\s|\over v_0^2}\right)^{1/4} +
\left({\l^2\over v_0}\right)^N \right]
\;,\Eq(5.27)$$
which immediately implies the bound in the second line of \equ(2.19c). \qed

\*

\sub(5.8) {\it Proof of Lemma \secc(2.16).}
The Hessian matrix $\tilde M$, defined in \equ(1.15), is a real matrix;
hence, we have to show that
$$ \sum_{n,m=-[Q/2]}^{[(Q-1)/2]} x_n \tilde M_{nm} x_m
> 0 \; , \Eq(5.28) $$
for any $\{x_n\}_{n=-[Q/2]}^{[(Q-1)/2]}\in\RRR^{Q-3}$.
This will be done by writing
$$ \eqalign{
& \sum_{n,m=-[Q/2]}^{[(Q-1)/2]} x_n \tilde M_{nm} x_m \ge \cr
& \qquad \sum_{n=-[Q/2]}^{[(Q-1)/2]} x_n^2 \left[ \tilde M_{nn}-
{1\over 2}\sum_{m=-[Q/2] \atop m\not=n}^{[(Q-1)/2]}
\left(|\tilde M_{nm}|+|\tilde M_{mn}| \right) \right]
\; , \cr} \Eq(5.29) $$
and showing that the right hand side of the above equation is strictly
positive.

Let us find first a lower bound for $\tilde M_{nn}$.
If $|n|\not=1$, by \equ(1.15), \equ(1.12) and \equ(2.1) we have
$$ \tilde M_{nn} = 1+\l^2 c_n(\s)-
\l^2 {\dpr \tilde\r_n\over \partial \F_n} \; , \Eq(5.30) $$
where $1+\l^2 c_n(\s)\ge 1/2$, see \sec(2.12), and
$\partial \tilde\r_n/ \dpr\F_n$ obeys to the same bound of
$\partial \tilde\r_n/\partial\s$, see \sec(5.3),
up to the factor $|n|^{-N}$:
simply note that the derivatives can act only on the vertex functions
(and not on the propagators),
and $|\dpr\F_n/\dpr\s|\le|n|^{-N}$ has to be replaced with
$|\dpr\F_n/\dpr\F_n|\le 1$. Then, analogously to \equ(5.11a),
we obtain, for any $q>0$,
$$ \left| \l^2 {\dpr \tilde\r_n^q\over \partial \F_n} \right|
\le \l^2 D v_0 \left( {C\over v_0^2} \right)^q  q^2 (3q)^N (|\s|Q)^{[q/2]}
\; , \Eq(5.30a) $$
so that, if $|\s|Qv_0^{-4}$ is small enough, we have
$$ \left| \l^2{\partial \tilde\r_n \over
\dpr\F_n} \right| \le C \l^2 v_0^{-1} 3^N N! \;.\Eq(5.30b) $$
It follows that
$$ \tilde M_{nn} \ge {1\over 3} \; , \qquad |n|\neq1 \; , \Eq(5.31) $$
for $\l$ satisfying \equ(1.16), with $\e$ small enough and $K\ge 3$.

In the case $n=1$ (the case $n=-1$ is discussed in the same way) we have
$$ \tilde M_{11}=1+\l^2 c_1(\s)+\l\s{\partial c_1(\s)
\over\dpr\hat\f_1}-\l{\partial \tilde\r_1\over\partial\hat\f_1} \;.\Eq(5.32) $$
Note that our definitions of $c_1(\s)$ and $\tilde\r_n(\s,\F)$ do not
distinguish the dependence on $\hat\f_1$ and $\hat\f_{-1}$, which are equal
in the fixed points we are studying (see discussion in \sec(1.5)).
However, in the definition of $\tilde M$, $\hat\f_1$ and $\hat\f_{-1}$
have to be treated as independent variables. By taking into account this
remark and by using Lemmata \secc(2.13a) and \secc(2.13b),
with $\tilde\e$ and $|\s|v_0^{-2}$ small enough, we get
$$ \eqalign{
\l\s{\partial c_1(\s)\over\dpr\hat\f_1}&=
{1\over 2}\l^2\s {\partial c_1(\s)\over\dpr \s} \ge
{1\over 6\p} {\l^2\over v_0} \; , \cr
\big| 1+\l^2 c_1(\s) \big| & = \Big| { \l^2 \tilde\r_1\over \s} \Big|
\le {C\l^2\over v_0} \left[ \left( {|\s|\over v_0^2} \right)^{1/4} +
\left({\l^2\over v_0}\right)^N \right] \; , \cr
\left|\l{\partial \tilde\r_1\over\partial\hat\f_1}\right|&\le
\Big| \l^2 {\partial \tilde\r_1\over\partial \s} \Big|
\le {C\l^2\over v_0} \left[\left({|\s|\over v_0^2}\right)^{1/4} +
\left( {\l^2\over v_0} \right)^N \right] \; , \cr} \Eq(5.33) $$
so that
$$|\tilde M_{nn}| \ge {1\over 8\p} {\l^2\over v_0} \; ,
\qquad n=\pm 1 \; , \Eq(5.34) $$
under the hypotheses of Theorem \secc(1.7).

The non diagonal terms ($n\neq m$) are of the form
$$\tilde M_{nm}=\l^2\F_m {\partial c_m(\s)\over
\partial\s}{\d_{n,1}+\d_{n,-1}\over 2}-\l{\partial
\tilde\r_m\over\partial\hat\f_n} \;.\Eq(5.35)$$
By using \equ(2.6a) and \equ(2.12), the first term in the r.h.s. of
\equ(5.35), where $m\not=n$ implies $|m|>1$, can be bounded as
$$\left| \l^2 \F_m {\partial c_m(\s)\over \partial\s}{\d_{n,1}+\d_{n,-1}\over 2}
\right| \le C\left( {\l^2\over v_0} \right)^{N+1}\;.\Eq(5.35a)$$
Moreover, by proceeding again as in \S 5.3, we obtain
$$ \left| \l{\dpr \tilde\r_m^q\over \partial \hat\f_n} \right|
\le \l^2 D v_0 \left( {C\over v_0^2} \right)^q  q^2
(3q)^N (|\s|Q)^{[q/2]} \; , \Eq(5.35b) $$
so that
$$\sum_{n=-[Q/2];n\not=m}^{[(Q-1)/2]}\sum_{q=2}^\io \left| \l{\dpr
\tilde\r_m^q\over\partial \hat\f_n} \right| \le
C\l^2 3^N N! {|\s|Q^2\over v_0^5}\; .\Eq(5.36)$$

The contributions with $q=1$ need an improved bound. Let us first suppose that
$|n|>1$; in this case the derivative can act only on the vertex function
of the graphs contributing to $\tilde\r_m^1$. Then, if the derivative is
different from $0$, the vertex function is equal to $\F_n$ and,
since $m-n\not=0$, at least one of the two propagators must have different
$\o$ indices; this follows from \equ(4.2), which also implies that the integer
which multiplies $\pp_F$ in the value \equ(4.1) of the graph is different from
$0$. Hence, by using \equ(4.5), \equ(4.51) and the fact that $|n-m|\le 2$,
we get
$$\sum_{n=-[Q/2];n\not=m}^{[(Q-1)/2]}\left| \l^2{\partial
\tilde\r_m^1\over\partial \F_n} \right| \le
C\l^2 {|\s|Q\over v_0^3} \left(1+\log{v_0^2\over |\s|}\right)\; .\Eq(5.37)$$
The case $q=1$ and $|n|=1$ can be treated in a similar way; the main
difference is that the derivative can act also on the propagators of the
graphs contributing to $\tilde\r_m^1$, but it is still true that the integer
which multiplies $\pp_F$ in the value \equ(4.1) of each graph is different
from $0$, an essential point in the previous bound, since it allowed to use
the improved bound \equ(4.51) in place of \equ(4.7). By using again \equ(4.5)
and \equ(4.51), as well as the improved bounds (with respect to \equ(4.7c))
\equ(4.7d) and \equ(4.7h), we get again the bound \equ(5.37).

The r.h.s. of \equ(5.35a), \equ(5.36) and \equ(5.37) can be
made arbitrarily small with respect to $\l^2/v_0$, by suitably choosing
the constants in \equ(1.16); hence Lemma \secc(2.16) is proved. \qed

\pagina

\centerline{\titolo References}
\*
\halign{\hbox to 1.2truecm {[#]\hss} &
        \vtop{\advance\hsize by -1.25 truecm \0#}\cr

AL& {S. Aubry, P.Y. Le Daeron:
The discrete FrenkelKontorova model and its extensions.
{\it Phys. D} {\bf 8}, 381--422 (1983). }\cr
AAR& {S. Aubry, G. Abramovici, J. Raimbault:
Chaotic polaronic and bipolaronic
states in the adiabatic Holstein model.
{\it J. Stat. Phys.} {\bf  67}, 675--780 (1992). }\cr
BGM& {G. Benfatto, G. Gentile, V. Mastropietro:
Electrons in a lattice with an incommensurate potential.
{\it J. Stat. Physics} {\bf 89}, 655--708 (1997). }\cr
BM& {C. Baesens, R.S. MacKay:
Improved proof of existence of chaotic polaronic and bipolaronic
states for the adiabatic Holstein model and generalizations.
{\it Nonlinearity} {\bf  7}, 59--84 (1994). }\cr
D& {H. Davenport:
{\sl The Higher Arithmetic}, Dover, New York, 1983.}\cr
F& {H. Fr\"ohlich:
On the theory of superconductivity: the one-dimensional case.
{\it Proc. Roy. Soc. A} {\bf 223}, 296--305 (1954). }\cr
FGM& {J.K. Freericks, Ch. Gruber, N. Macris:
Phase separation in the binary-alloy problem: The one-dimensional spinless
Falicov-Kimball model.
{\it Phys. Rev. B} {\bf 53}, 16189--16196 (1996).}\cr
H& {T. Holstein: Studies of polaron motion, part 1.
The molecular-crystal model.
{\it Ann Phys.} {\bf 8}, 325--342 (1959). }\cr
KL& {T. Kennedy, E.H. Lieb:
Proof of the Peierls instability in one dimension.
{\it Phys. Rev. Lett.} {\bf 59}, 1309--1312 (1987). }\cr
LM& {J.L. Lebowitz, N. Macris:
Low-temperature phases of itinerant Fer\-mions interacting with
classical phonons: the static Holstein model.
{\it J. Stat. Phys.} {\bf 76}, 91--123, (1994). }\cr
LRA& {P.A. Lee, T.M. Rice, P.W. Anderson:
Conductivity from Charge or Spin Density Waves.
{\it Solid State Comm.} {\bf 14}, 703--720 (1974). }\cr
P& {R.E. Peierls:
{\sl Quantum theory of solids}.
Clarendon, Oxford, 1955. }\cr
}
%

\bye